\documentclass[aps,prl,twocolumn,showpacs,superscriptaddress,groupedaddress]{revtex4}

\usepackage{color}
\usepackage{graphicx}
\usepackage{dcolumn}
\usepackage{bm}
\usepackage{float}
\usepackage{amssymb}
\usepackage{epsfig}
\usepackage{subfigure}
\usepackage{rotating}
\usepackage{placeins}
\usepackage{hyperref}

\newcommand{\infb}{fb$^{-1}$}

\newcommand{\madgraph}{{\sc madgraph}}

\newcommand{\pythia} {{\sc pythia}}
\newcommand{\alpgen} {{\sc alpgen}}

\newcommand{\SingleTop}{{\sc singletop}}

\newcommand{\met}       {\mbox{$\not\!\!E_T$}}

\newcommand{\bbbar}{b\bar{b}}

\newcommand{\jp}{$J^{P}$}
\newcommand{\jpzp}{$J^{P}=0^{+}$}
\newcommand{\jpzm}{$J^{P}=0^{-}$}
\newcommand{\jptp}{$J^{P}=2^{+}$}

\newcommand{\whl}{$WH\rightarrow \ell\nu b\bar{b}$}
\newcommand{\zhl}{$ZH\rightarrow \ell\ell b\bar{b}$}
\newcommand{\zhv}{$ZH\rightarrow \nu\nu b\bar{b}$}
\newcommand{\hbb}{$H\rightarrow b\bar{b}$}
\newcommand{\llbb}{$\ell\ell b\bar{b}$}
\newcommand{\lvbb}{$\ell\nu b\bar{b}$}
\newcommand{\vvbb}{$\nu\nu b\bar{b}$}
\newcommand{\vbb}{$VH\rightarrow Vb\bar{b}$}

\newcommand{\clzp}{$CL_{0^{+}}$}
\newcommand{\clzm}{$CL_{0^{-}}$}
\newcommand{\cltp}{$CL_{2^{+}}$}
\newcommand{\ftplus}{f_{2^{+}}}
\newcommand{\fzminus}{f_{0^{-}}}
\newcommand{\fzminusm}{$f_{0^{-}}$}
\newcommand{\ftplusm}{$f_{2^{+}}$}
\newcolumntype{q}[1]{{.}{.}{#1}}
\newcommand\T{\rule{0pt}{2.6ex}}       
\newcommand\B{\rule[-1.2ex]{0pt}{0pt}}

\begin{document}

\hspace{5.2in}\mbox{FERMILAB-PUB-14-242-E}

\title{Constraints on models for the Higgs boson with exotic \\
spin and parity in $\boldsymbol{VH\rightarrow Vb\bar{b}}$ final states}
\affiliation{LAFEX, Centro Brasileiro de Pesquisas F\'{i}sicas, Rio de Janeiro, Brazil}
\affiliation{Universidade do Estado do Rio de Janeiro, Rio de Janeiro, Brazil}
\affiliation{Universidade Federal do ABC, Santo Andr\'e, Brazil}
\affiliation{University of Science and Technology of China, Hefei, People's Republic of China}
\affiliation{Universidad de los Andes, Bogot\'a, Colombia}
\affiliation{Charles University, Faculty of Mathematics and Physics, Center for Particle Physics, Prague, Czech Republic}
\affiliation{Czech Technical University in Prague, Prague, Czech Republic}
\affiliation{Institute of Physics, Academy of Sciences of the Czech Republic, Prague, Czech Republic}
\affiliation{Universidad San Francisco de Quito, Quito, Ecuador}
\affiliation{LPC, Universit\'e Blaise Pascal, CNRS/IN2P3, Clermont, France}
\affiliation{LPSC, Universit\'e Joseph Fourier Grenoble 1, CNRS/IN2P3, Institut National Polytechnique de Grenoble, Grenoble, France}
\affiliation{CPPM, Aix-Marseille Universit\'e, CNRS/IN2P3, Marseille, France}
\affiliation{LAL, Universit\'e Paris-Sud, CNRS/IN2P3, Orsay, France}
\affiliation{LPNHE, Universit\'es Paris VI and VII, CNRS/IN2P3, Paris, France}
\affiliation{CEA, Irfu, SPP, Saclay, France}
\affiliation{IPHC, Universit\'e de Strasbourg, CNRS/IN2P3, Strasbourg, France}
\affiliation{IPNL, Universit\'e Lyon 1, CNRS/IN2P3, Villeurbanne, France and Universit\'e de Lyon, Lyon, France}
\affiliation{III. Physikalisches Institut A, RWTH Aachen University, Aachen, Germany}
\affiliation{Physikalisches Institut, Universit\"at Freiburg, Freiburg, Germany}
\affiliation{II. Physikalisches Institut, Georg-August-Universit\"at G\"ottingen, G\"ottingen, Germany}
\affiliation{Institut f\"ur Physik, Universit\"at Mainz, Mainz, Germany}
\affiliation{Ludwig-Maximilians-Universit\"at M\"unchen, M\"unchen, Germany}
\affiliation{Panjab University, Chandigarh, India}
\affiliation{Delhi University, Delhi, India}
\affiliation{Tata Institute of Fundamental Research, Mumbai, India}
\affiliation{University College Dublin, Dublin, Ireland}
\affiliation{Korea Detector Laboratory, Korea University, Seoul, Korea}
\affiliation{CINVESTAV, Mexico City, Mexico}
\affiliation{Nikhef, Science Park, Amsterdam, the Netherlands}
\affiliation{Radboud University Nijmegen, Nijmegen, the Netherlands}
\affiliation{Joint Institute for Nuclear Research, Dubna, Russia}
\affiliation{Institute for Theoretical and Experimental Physics, Moscow, Russia}
\affiliation{Moscow State University, Moscow, Russia}
\affiliation{Institute for High Energy Physics, Protvino, Russia}
\affiliation{Petersburg Nuclear Physics Institute, St. Petersburg, Russia}
\affiliation{Instituci\'{o} Catalana de Recerca i Estudis Avan\c{c}ats (ICREA) and Institut de F\'{i}sica d'Altes Energies (IFAE), Barcelona, Spain}
\affiliation{Uppsala University, Uppsala, Sweden}
\affiliation{Taras Shevchenko National University of Kyiv, Kiev, Ukraine}
\affiliation{Lancaster University, Lancaster LA1 4YB, United Kingdom}
\affiliation{Imperial College London, London SW7 2AZ, United Kingdom}
\affiliation{The University of Manchester, Manchester M13 9PL, United Kingdom}
\affiliation{University of Arizona, Tucson, Arizona 85721, USA}
\affiliation{University of California Riverside, Riverside, California 92521, USA}
\affiliation{Florida State University, Tallahassee, Florida 32306, USA}
\affiliation{Fermi National Accelerator Laboratory, Batavia, Illinois 60510, USA}
\affiliation{University of Illinois at Chicago, Chicago, Illinois 60607, USA}
\affiliation{Northern Illinois University, DeKalb, Illinois 60115, USA}
\affiliation{Northwestern University, Evanston, Illinois 60208, USA}
\affiliation{Indiana University, Bloomington, Indiana 47405, USA}
\affiliation{Purdue University Calumet, Hammond, Indiana 46323, USA}
\affiliation{University of Notre Dame, Notre Dame, Indiana 46556, USA}
\affiliation{Iowa State University, Ames, Iowa 50011, USA}
\affiliation{University of Kansas, Lawrence, Kansas 66045, USA}
\affiliation{Louisiana Tech University, Ruston, Louisiana 71272, USA}
\affiliation{Northeastern University, Boston, Massachusetts 02115, USA}
\affiliation{University of Michigan, Ann Arbor, Michigan 48109, USA}
\affiliation{Michigan State University, East Lansing, Michigan 48824, USA}
\affiliation{University of Mississippi, University, Mississippi 38677, USA}
\affiliation{University of Nebraska, Lincoln, Nebraska 68588, USA}
\affiliation{Rutgers University, Piscataway, New Jersey 08855, USA}
\affiliation{Princeton University, Princeton, New Jersey 08544, USA}
\affiliation{State University of New York, Buffalo, New York 14260, USA}
\affiliation{University of Rochester, Rochester, New York 14627, USA}
\affiliation{State University of New York, Stony Brook, New York 11794, USA}
\affiliation{Brookhaven National Laboratory, Upton, New York 11973, USA}
\affiliation{Langston University, Langston, Oklahoma 73050, USA}
\affiliation{University of Oklahoma, Norman, Oklahoma 73019, USA}
\affiliation{Oklahoma State University, Stillwater, Oklahoma 74078, USA}
\affiliation{Brown University, Providence, Rhode Island 02912, USA}
\affiliation{University of Texas, Arlington, Texas 76019, USA}
\affiliation{Southern Methodist University, Dallas, Texas 75275, USA}
\affiliation{Rice University, Houston, Texas 77005, USA}
\affiliation{University of Virginia, Charlottesville, Virginia 22904, USA}
\affiliation{University of Washington, Seattle, Washington 98195, USA}
\author{V.M.~Abazov} \affiliation{Joint Institute for Nuclear Research, Dubna, Russia}
\author{B.~Abbott} \affiliation{University of Oklahoma, Norman, Oklahoma 73019, USA}
\author{B.S.~Acharya} \affiliation{Tata Institute of Fundamental Research, Mumbai, India}
\author{M.~Adams} \affiliation{University of Illinois at Chicago, Chicago, Illinois 60607, USA}
\author{T.~Adams} \affiliation{Florida State University, Tallahassee, Florida 32306, USA}
\author{J.P.~Agnew} \affiliation{The University of Manchester, Manchester M13 9PL, United Kingdom}
\author{G.D.~Alexeev} \affiliation{Joint Institute for Nuclear Research, Dubna, Russia}
\author{G.~Alkhazov} \affiliation{Petersburg Nuclear Physics Institute, St. Petersburg, Russia}
\author{A.~Alton$^{a}$} \affiliation{University of Michigan, Ann Arbor, Michigan 48109, USA}
\author{A.~Askew} \affiliation{Florida State University, Tallahassee, Florida 32306, USA}
% ...
%
%
\collaboration{The D0 Collaboration\footnote{with visitors from
$^{a}$Augustana College, Sioux Falls, SD, USA,
$^{b}$The University of Liverpool, Liverpool, UK,
$^{c}$DESY, Hamburg, Germany,
$^{d}$Universidad Michoacana de San Nicolas de Hidalgo, Morelia, Mexico
$^{e}$SLAC, Menlo Park, CA, USA,
$^{f}$University College London, London, UK,
$^{g}$Centro de Investigacion en Computacion - IPN, Mexico City, Mexico,
$^{h}$Universidade Estadual Paulista, S\~ao Paulo, Brazil,
$^{i}$Karlsruher Institut f\"ur Technologie (KIT) - Steinbuch Centre for Computing (SCC),
D-76128 Karlsruhe, Germany,
$^{j}$Office of Science, U.S. Department of Energy, Washington, D.C. 20585, USA,
$^{k}$American Association for the Advancement of Science, Washington, D.C. 20005, USA,
$^{l}$Kiev Institute for Nuclear Research, Kiev, Ukraine
and
$^{m}$University of Maryland, College Park, Maryland 20742, USA.
}} \noaffiliation
\vskip 0.25cm

\date{August 28, 2014}

\begin{abstract}
We present constraints on models containing non-standard model values
for the spin $J$ and parity $P$ of the Higgs boson, $H$, in up to
9.7~fb$^{-1}$ of $p\bar{p}$ collisions at $\sqrt{s} = $ 1.96~TeV
collected with the D0 detector at the Fermilab Tevatron
Collider. These are the first studies of Higgs boson $J^{P}$ with
fermions in the final state. In the $ZH\rightarrow \ell\ell b\bar{b}$, $WH\rightarrow
\ell\nu b\bar{b}$, and $ZH\rightarrow \nu\nu b\bar{b}$ final states,
we compare the standard model (SM) Higgs boson prediction,
$J^{P}=0^{+}$, with two alternative hypotheses, $J^{P}=0^{-}$ and
$J^{P}=2^{+}$.  We use a likelihood ratio to quantify the degree to
which our data are incompatible with non-SM
$J^{P}$ predictions for a range of possible production rates. 
Assuming that the production rate in the signal models considered is
equal to the SM prediction, we reject the $J^{P}=0^{-}$ and
$J^{P}=2^{+}$ hypotheses at the 97.6$\%$ CL and at the 99.0$\%$ CL,
respectively. The expected exclusion sensitivity for a $J^{P}=0^{-}$
($J^{P}=2^{+}$) state is at the 99.86$\%$ (99.94$\%$) CL. Under the
hypothesis that our data is the result of a combination of the SM-like
Higgs boson and either a $J^{P}=0^{-}$ or a $J^{P}=2^{+}$ signal, we
exclude a $J^{P}=0^{-}$ fraction above 0.80 and a $J^{P}=2^{+}$
fraction above 0.67 at the 95$\%$ CL. The expected exclusion covers
$J^{P}=0^{-}$ ($J^{P}=2^{+}$) fractions above 0.54 (0.47).
\end{abstract}

\maketitle

After the discovery of a Higgs boson, $H$, at the CERN Large Hadron
Collider (LHC)~\cite{atlas-obs,cms-obs} in bosonic final states, and
evidence for its decay to to a pair of $b$ quarks at the Tevatron experiments~\cite{Aaltonen:2012qt},
and to pairs of fermions at the CMS experiment~\cite{Chatrchyan:2014tb},
it is important to determine
the new particle's properties using all decay modes available. In
particular, the spin and parity of the Higgs boson are important in
determining the framework of the mass generation mechanism. The SM
predicts that the Higgs boson is a CP-even spin-0 particle (\jpzp). If
the Higgs boson is indeed a single boson, the observation of its decay
to two photons at the LHC precludes spin~1 according to the
Landau-Yang theorem~\cite{Landau:1948kw,Yang:1950rg}. Other \jp\
possibilities are possible. An admixture of \jpzp\ and \jpzm\
can arise in Two-Higgs-Doublet models
(2HDM)~\cite{2hdm:Lee,Cervero2012255} of type II such as found in
supersymmetric models. A boson with tensor couplings (\jptp) can arise
in models with extra dimensions~\cite{Fok:2012zk}.  The ATLAS and CMS
Collaborations have examined the possibility that the $H$ boson has
\jpzm\ or \jptp\ using its decays to $\gamma\gamma$, $ZZ$, and $WW$
states~\cite{Chatrchyan:2012jja,Aad:2013xqa,Chatrchyan:2013iaa,Chatrchyan:2013mxa,Chatrchyan:2014obd}. 
The \jpzm\ hypothesis is excluded at the 97.8\%\
and 99.95\%\ CL by the ATLAS and CMS Collaborations, respectively, in
the $H \rightarrow ZZ \rightarrow 4\ell$ decay mode. Likewise, the
\jptp\ hypothesis is excluded at the $\geq 99.9$\%\ CL by the ATLAS
Collaboration when combining all bosonic decay modes, and at the $\geq
97.7$\%\ CL by the CMS Collaboration in the $H \rightarrow ZZ
\rightarrow 4\ell$ decay mode (depending on the production processes and the 
quark-mediated fraction of the production processes). However, the
\jp\ character of Higgs bosons decaying to pairs of fermions, and in
particular to $b\bar{b}$, has not yet been studied. In this Letter we
present tests of non-SM models describing production of bosons with a
mass of 125~GeV, \jpzm\ or \jptp, and decaying to $b\bar{b}$. We
explore two scenarios for each of the hypotheses: (a) the new boson is
a \jpzm\ (\jptp) particle and (b) the observed resonance is either a
combination of these non-SM \jp\ states and a \jpzp\ state or distinct
states with degenerate mass. In the latter case, we do not consider
interference effects between states.

Unlike the LHC \jp\ measurements, our ability to distinguish different
Higgs boson \jp\ assignments is not based primarily on the angular
analysis of the Higgs boson decay products. It is instead based on the
kinematic correlations between the vector boson $V (V = W, Z)$ and the
Higgs boson in$\,$\textit{VH} associated production. Searches for
associated$\,$\textit{VH} production are sensitive to the different
kinematics of the various \jp\ combinations in several observables,
especially the invariant mass of the$\,$\textit{VH} system, due to
the dominant $p$ and $d$ wave contributions to the \jpzm\ and \jptp\
production
processes~\cite{Ellis:2012xd,Ellis:2013ywa,Miller:2001beta}.  The
$p$ and $d$ wave contributions to the production cross
sections near threshold vary as $\beta^{3}$ and $\beta^{5}$,
respectively, whereas the $s$ wave contribution for the SM
Higgs boson varies as $\beta$, where $\beta$ is 
the ratio of the Higgs boson momentum and energy.

To test compatibility of non-SM \jp\ models with data we use the D0 studies of 
\zhl~\cite{Abazov:2013mla}, \whl~\cite{Abazov:lvjets}, 
and \zhv~\cite{Abazov:2012hv} with no modifications to the event
selections. Lepton flavors considered in the \whl\ and \zhl\ analyses
include electrons and muons. Events with taus that decay to these
leptons are considered as well, although their contribution is
small. The D0 detector is described in
Refs.~\cite{Abazov:2005pn,Abolins:2007yz,Angstadt:2009ie}. 

We use 9.5--9.7~\infb\ of integrated luminosity collected with the D0
detector satisfying relevant data-quality requirements in each of the
three analyses. The SM background processes are either estimated from
dedicated data samples (multijet backgrounds), or from Monte Carlo
(MC) simulation.  The $V$+jets and $t\bar{t}$ processes are generated
using \alpgen~\cite{Mangano:2002ea}, single top processes are
generated using \SingleTop~\cite{Boos:2006af}, and diboson
($\!$\textit{VV}) processes are generated using
\pythia~\cite{Sjostrand:2006za}.  The SM Higgs boson processes are
also generated using \pythia. The signal samples for the \jpzm\ and
\jptp\ hypotheses are generated using
\madgraph~5~\cite{Alwall:2011uj}. We have verified that \jpzp\ samples produced with
\madgraph\ agree well with the SM \pythia\ prediction. 

In the following, we denote a non-SM Higgs boson as $X$, reserving the
label $H$ for the SM \jpzp\ Higgs boson.
\madgraph\ can simulate several non-SM models, as well as 
user-defined models. These new states are introduced via dimension-5
Lagrangian operators~\cite{Ellis:2013ywa}.  The \jpzm\ samples are
created using a model from the authors of
Ref.~\cite{Ellis:2012xd}. The non-SM Lagrangian can be expressed
as~\cite{Ellis:2013ywa}
$\mathcal{L}_{0^{-}}=\frac{c^{A}_{V}}{\Lambda}AF_{\mu\nu}\tilde{F}^{\mu\nu},$
where $F_{\mu\nu}$ is the field-strength tensor for the vector boson,
$A$ is the new boson field, $c^{A}_{V}$ is a coupling term, and
$\Lambda$ is the scale at which new physics effects arise. The \jptp\
signal samples are created using a Randall-Sundrum (RS) model, an
extra-dimension model with a massive \jptp\ particle that has
graviton-like couplings~\cite{RS:Oct1999, RS:Dec1999, MGRS:2008,
MGRS:2011}. This model's Lagrangian can be expressed as
$\mathcal{L}_{2^{+}}=\frac{c^{G}_{V}}{\Lambda}G^{\mu\nu}T_{\mu\nu},$
where $G^{\mu\nu}$ represents the \jptp\ particle, $c^{G}_{V}$ is a
coupling term, $T_{\mu\nu}$ is the stress-energy tensor of the vector
boson, and $\Lambda$ is the effective Planck mass~\cite{Fok:2012zk}.
The mass of the non-SM Higgs-like particle $X$ is set to 125~GeV, a
value close to the mass measured by the LHC
Collaborations~\cite{atlas-obs,cms-obs} and also consistent with
measurements at the Tevatron~\cite{Aaltonen:2012qt}. We study the
decay of $X$ to $b\bar{b}$ only. For our initial sample normalization
we assume that the ratio $\mu$ of the product of the cross section and the branching
fraction, $\sigma(VX)\times{\cal B}(X\to\bbbar)$, to the SM prediction 
is $\mu=1.0$~\cite{Baglio:2010um,Brein:2012jc}, and subsequently 
define exclusion regions as functions of $\mu$. We use
the CTEQ6L1 PDF set for sample generation, and \pythia\ for parton
showering and hadronization.  The MC samples are processed by the full
D0 detector simulation. To reproduce the effect of multiple $p\bar{p}$
interactions in the same beam crossing, each simulated event is
overlaid with an event from a sample of random beam crossings with the
same instantaneous luminosity profile as the data. The events are then
reconstructed with the same programs as the data.

All three analyses employ a $b$-tagging algorithm based on track
impact parameters, secondary vertices, and event topology to select
jets that are consistent with originating from a $b$
quark~\cite{Abazov:2013gaa,Abazov:2010ab}.

The \zhl\ analysis~\cite{Abazov:2013mla} selects events with two
isolated charged leptons and at least two jets. A kinematic fit
corrects the measured jet energies to their best fit values based on
the constraints that the dilepton invariant mass should be consistent
with the $Z$ boson mass~\cite{Beringer:1900zz} and that the total
transverse momentum of the leptons and jets should be consistent with
zero. The event sample is further divided into orthogonal
``single-tag'' (ST) and ``double-tag'' (DT) channels according to the
number of $b$-tagged jets. The SM Higgs boson search uses random
forest (RF)~\cite{Hocker:2007ht} discriminants to provide
distributions for the final statistical analysis. The first RF is
designed to discriminate against $t\bar{t}$ events and divides events
into $t\bar{t}$-enriched and $t\bar{t}$-depleted ST and DT regions. In
this study only events in the $t\bar{t}$-depleted ST and DT regions
are considered. These regions contain $\approx94\%$ of the SM Higgs
signal.

The \whl\ analysis~\cite{Abazov:lvjets} selects events with one
charged lepton, significant imbalance in the transverse energy (\met),
and two or three jets. This search is also sensitive to the \zhl\
process when one of the charged leptons is not identified. Using the
outputs of the $b$-tagging algorithm for all selected jets, events are
divided into four orthogonal $b$-tagging categories, ``one-tight-tag''
(1TT), ``two-loose-tag'' (2LT), ``two-medium-tag'' (2MT), and
``two-tight-tag'' (2TT). Looser $b$-tagging categories correspond to
higher efficiencies for true $b$ quarks and higher fake rates. Outputs
from boosted decision trees (BDTs)~\cite{Hocker:2007ht}, trained
separately for each jet multiplicity and tagging category, serve as
the final discriminants in the SM Higgs boson search.

The \zhv\ analysis~\cite{Abazov:2012hv} selects events with large
\met\ and exactly two jets.  This search is also sensitive to the $WH$
process when the charged lepton from the $W\to\ell\nu$ decay is not
identified. A dedicated BDT is used to provide rejection of the large
multijet background. Two orthogonal $b$-tagging channels, medium (MT),
and tight (TT), use the sum of the $b$-tagging
discriminants of the two selected jets.  BDT classifiers,
trained separately for the different $b$-tagging categories, provide
the final discriminants in the SM Higgs boson search.

These three analyses are among the inputs to the D0 SM Higgs boson
search~\cite{Abazov:2013zea}, yielding an excess above the SM
background expectation that is consistent both in shape and in
magnitude with a SM Higgs boson signal. The best fit to data for the
\hbb\ decay channel for the product of the signal cross section and
branching fraction, is $\mu=1.23^{+1.24}_{-1.17}$ for a mass of
125~GeV. When including data from both Tevatron experiments, the best
fit to data yields $\mu=1.59^{+0.69}_{-0.72}$~\cite{Aaltonen:2013kxa}.

Discrimination between the \jp\ values of non-SM and SM hypotheses is achieved by
using mass information of the$\,$\textit{VX}~system. For the
\llbb\ final state we use the invariant mass of the two leptons and 
either the two highest $b$-tagged jets (DT) or the $b$-tagged jet and
the highest $p_T$ non-tagged jet (ST) as the final discriminating
variable. For the final states that have neutrinos, the discriminating
variable is the transverse mass of the$\,$\textit{VX}~system which is
defined as
$M_{T}^{2}=(E_{T}^{V}+E_{T}^{X})^{2}-(\vec{p}_{T}^{V}+\vec{p}_{T}^{X})^{2}$
where the transverse momenta of the $Z$ and $W$ bosons are
$\vec{p}_{T}^{Z}=\vec{\not\!\!E_{T}}$ and
$\vec{p}_{T}^{W}=\vec{\not\!\!E_{T}}+\vec{p}_{T}^{\ell}$. For the
\lvbb\ final state the two jets can either be one $b$-tagged jet (1TT)
and the highest $p_{T}$ non-tagged jet, or the two $b$-tagged jets
from any of the other three $b$-tagging categories: 2LT, 2MT, or
2TT. 

To improve the discrimination between the non-SM signals and
backgrounds in the \llbb\ and \vvbb\ final states, we use the
invariant mass of the dijet system, $M_{jj}$, to select two regions
with different signal purities. Events with dijet masses in the range
$100\leq M_{jj}\leq 150$~GeV ($70 \leq M_{jj}< 150$~GeV) for \llbb\
(\vvbb) final states comprise the ``high-purity'' region (HP), while
the remaining events are in the ``low-purity region'' (LP). As a
result of the kinematic fit, the HP region for the \llbb\ final state
is narrower than that for the \vvbb\ final state, given the
correspondingly narrower dijet mass peak. For the \lvbb\ final state
we use the final BDT output ($\mathcal{D}$) of the SM Higgs boson
search\cite{Abazov:lvjets}. Since events with
$\mathcal{D}\leq 0$ provide negligible sensitivity to SM or non-SM
signals, we do not consider them further. We separate the remaining
events into two categories with different signal purities. The LP
category consists of events with $0\le \mathcal{D}
\leq 0.5$, and the HP category of events with $ \mathcal{D} > 0.5$.

\begin{figure*}[htpb]
\centering{
\includegraphics[width=0.32\textwidth]{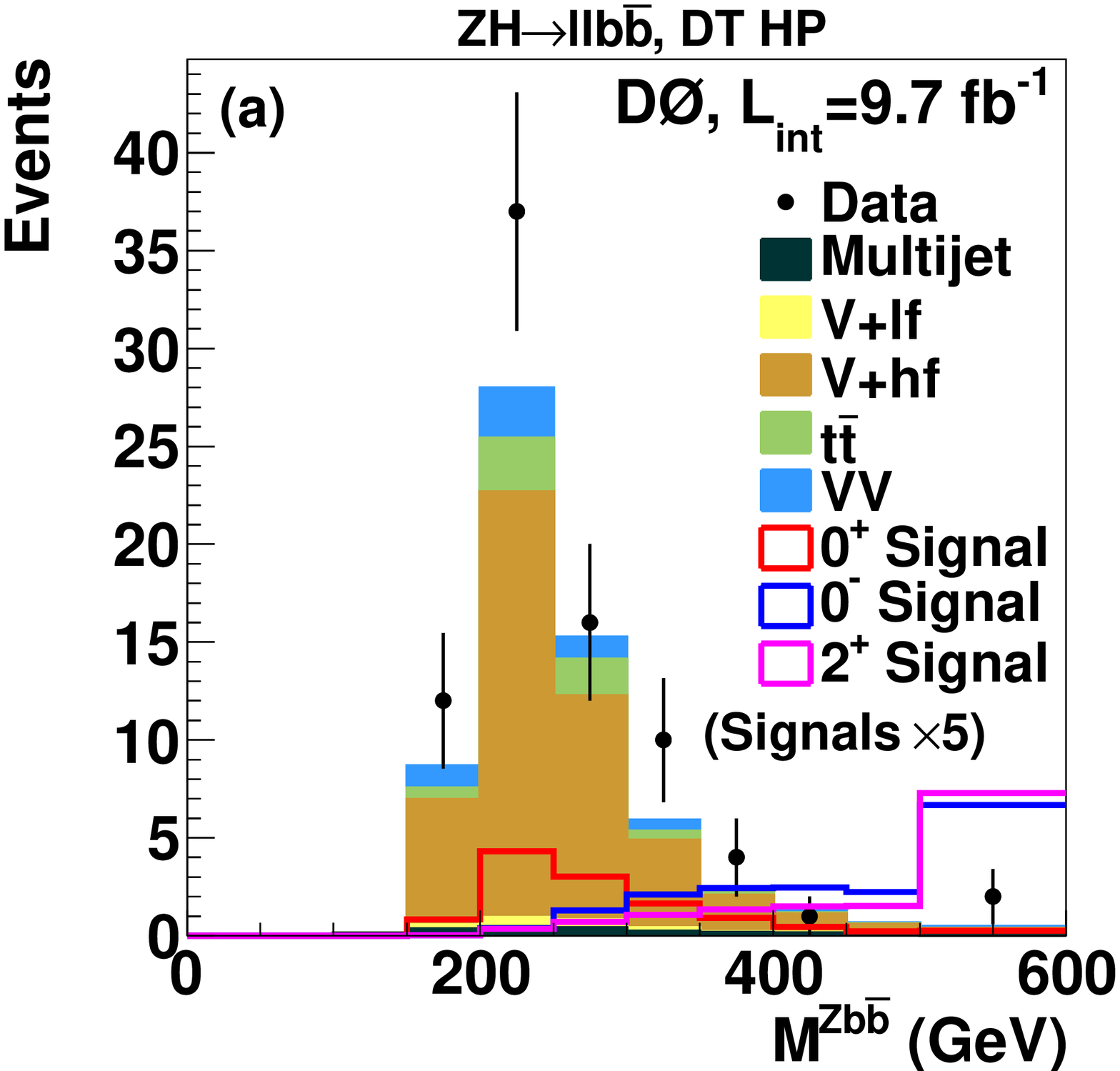}
\includegraphics[width=0.32\textwidth]{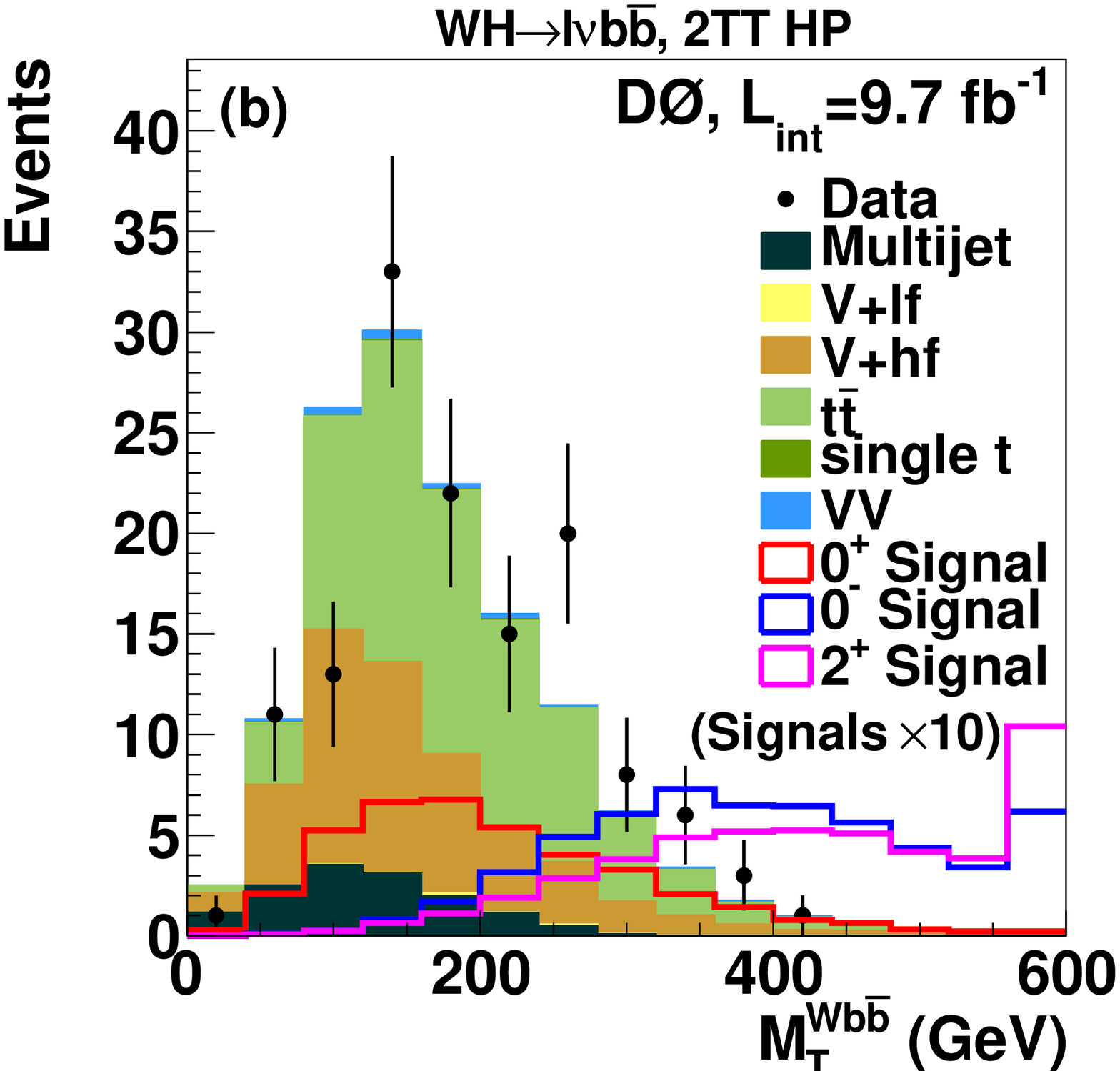}
\includegraphics[width=0.32\textwidth]{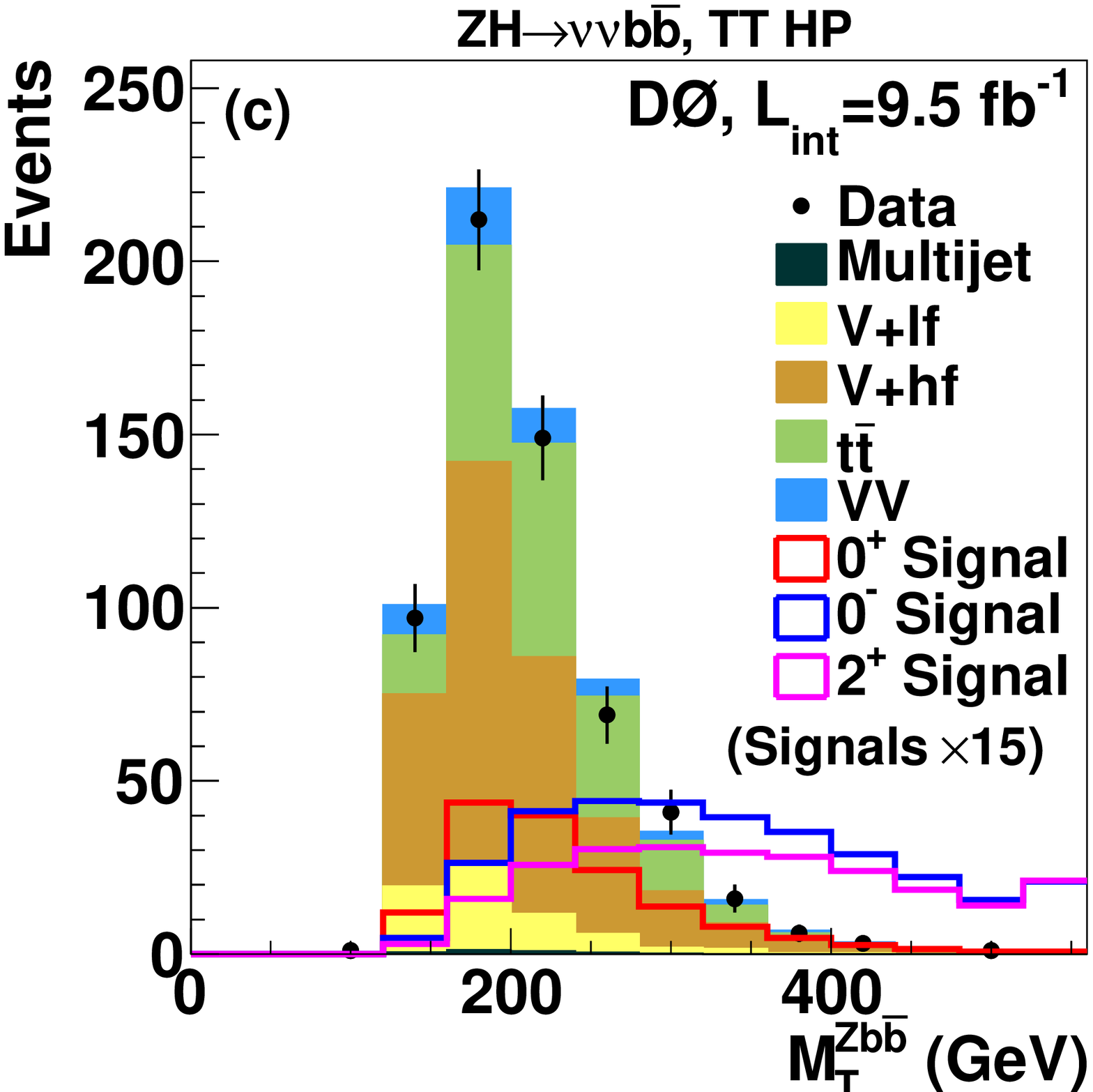}
}
\caption{(color online) (a) Invariant mass of the $\ell\ell b\bar{b}$ system in the \zhl\ high-purity double-tag (DT HP) channel, 
(b) transverse mass of the $\ell\nu b\bar{b}$ system in the \whl\
high-purity 2-tight-tag (2TT HP) channel, and (c) transverse mass of
the $\nu\nu b\bar{b}$ system in the \zhv\ high-purity tight-tag (TT
HP) channel. The \jpzm\ and \jptp\ samples are normalized to the
product of the SM cross section and branching fraction multiplied by
an additional factor. Heavy- and light-flavor quark jets are denoted
by lf and hf, respectively. Overflow events are included in the
highest mass bin. For all signals, a mass of 125~GeV for the $H$ or
$X$ boson is assumed.\label{fig:final_var}}
\end{figure*}

Figure~\ref{fig:final_var} illustrates the discriminating variables
for the three analysis channels in the high-purity categories for the
most sensitive $b$-tagging selections. Distributions for additional
subchannels can be found in Ref.~\cite{epaps}.

We perform the statistical analysis using a modified frequentist
approach~\cite{Read:2002hq,wade_tm,Abazov:2013zea}. We use a negative
log-likelihood ratio (LLR) as the test statistic for two hypotheses:
the null hypothesis, $H_0$, and the test hypothesis, $H_1$.  This LLR
is given by ${\rm LLR} = -2\ln\left(L_{H_1}/L_{H_0}\right)\label{eq:llr}$, 
where $L_{H_{x}}$ is the joint likelihood for hypothesis $x$ evaluated
over the number of bins in the final discriminating variable
distribution in each channel. To decrease the effect of systematic
uncertainties on the sensitivity, we fit the signals and backgrounds
by maximizing the likelihood functions by allowing the systematic
effects to vary within Gaussian constraints. This fit is performed
separately for both the $H_0$ and $H_1$ hypotheses for the data and
each pseudo-experiment.

We define $CL_s$ as~$CL_{H_1}/CL_{H_0}$ where $CL_{H_{x}}$ 
for a given hypothesis $H_x$ is~$CL_{H_{x}} = P_{H_{x}}({\rm LLR}\geq {\rm LLR}^{\rm obs})\label{eq:clx}$,
and LLR$^{\rm obs}$ is the LLR value observed in the data. $P_{H_{x}}$
is defined as the probability that the LLR falls beyond ${\rm
LLR}^{\rm obs}$ for the distribution of LLR populated by the $H_{x}$
model. For example, if $CL_s \le 0.05$ we exclude the $H_1$ hypothesis
in favor of the $H_0$ hypothesis at $\ge 95$\%\ CL.

Systematic uncertainties affecting both shape and rate are
considered. The systematic uncertainties for each individual analysis
are described in
Refs.~\cite{Abazov:2013mla,Abazov:lvjets,Abazov:2012hv}. A summary of
the major contributions follows. The largest contribution for all
analyses is from the uncertainties on the cross sections of the
simulated $V +$ heavy-flavor jets backgrounds which are
20\%--30\%. All other cross section uncertainties for simulated
backgrounds are less than 10\%. Since the multijet background is
estimated from data, its uncertainty depends on the size of the data
sample from which it is estimated, and ranges from 10\%\ to 30\%. All
simulated samples for the \whl\ and \zhv\ analyses have an uncertainty
of 6.1\%\ from the integrated luminosity~\cite{Andeen:2007zc}, whereas
the simulated samples from the \zhl\ analysis have uncertainties
ranging from 0.7\%--7\%\ arising from the fitted normalization to the
data~\cite{Abazov:2013mla}. All analyses take into account
uncertainties on the jet energy scale, resolution, and jet
identification efficiency for a combined uncertainty of $\approx7\%$.
The uncertainty on the $b$-tagging rate varies from 1\%--10\%\
depending on the number and quality of the tagged jets. The
correlations between the three analyses are described in
Ref.~\cite{Abazov:2013zea}.

\begin{figure}[t!]
\centering
\includegraphics[width=0.47\textwidth]{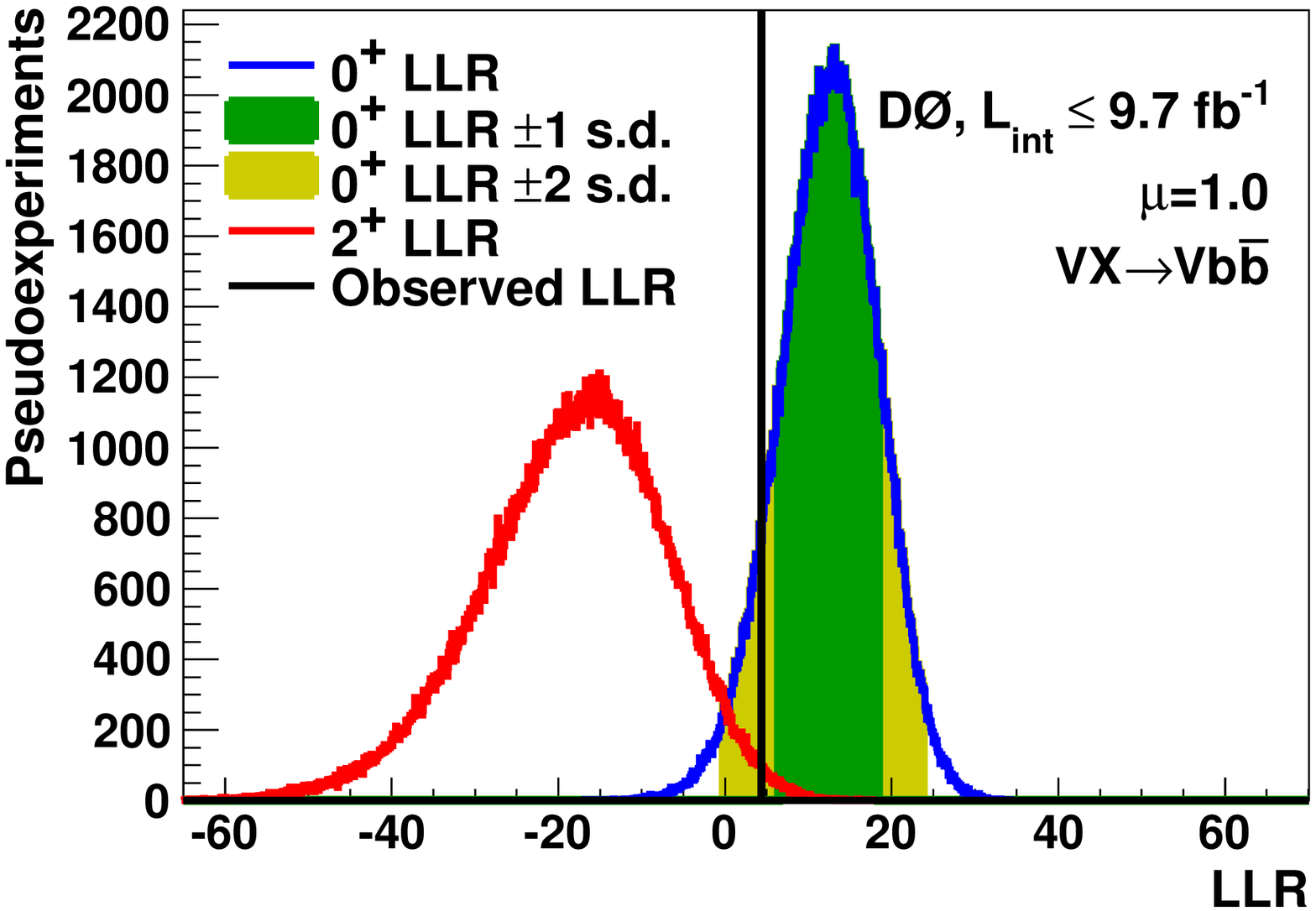}
\caption{(color online) LLR distributions comparing the \jpzp\ and the \jptp\ hypotheses for the combination. The \jpzp\ and \jptp\ samples are normalized 
to the product of the SM cross section and branching fraction. The
vertical solid line represents the observed LLR value assuming
$\mu=1.0$, while the dark and light shaded areas represent the 1 and 2
standard deviations (s.d.) on the expectation from the null hypothesis
$H_{0}$, respectively. \label{2pLLR}}
\end{figure}

In this Letter, the $H_0$ hypothesis always contains SM background
processes and the SM Higgs boson normalized to $\mu \times
\sigma^{SM}_{0^{+}}$. To test the non-SM cross section we assign the
$H_1$ hypothesis as the sum of the \jpzm\ or \jptp\ signal plus SM
background processes, with no contribution from the SM Higgs boson. We
calculate the $CL_{s}$ values using signal cross sections expressed as
$\mu \times \sigma^{SM}_{0^{+}}$ and evaluate the expected values for
each of these quantities by replacing $\rm{LLR}^{\rm obs}$ with
$\rm{LLR}_{0^{+}}^{\rm exp}$, the median expectation for the \jpzp\
hypothesis only. Figure~\ref{2pLLR} illustrates the LLR distributions
for the $H_{0}$ and \jptp\ $H_{1}$ hypotheses, and the observed LLR
value assuming $\mu=1.0$, a production rate compatible with both
Tevatron and LHC Higgs boson measurements. The similar plot for \jpzm\ 
is shown in Ref.~\cite{epaps}. We interpret $1-CL_{s}$ as
the confidence level at which we exclude the non-SM hypothesis for the
models considered in favor of the SM prediction of \jpzp\ for the
given value of $\mu$.  For $\mu=1.0$ we exclude the \jpzm\ (\jptp)
hypothesis at the 97.6\%\ (99.0\%) CL. The expected exclusions are at
the 99.86\%\ and 99.94\%\ CL. Results, including those for $\mu=1.23$, 
are given in Table~\ref{tab:sum}.

\begin{figure}[t!]
\centering
\includegraphics[width=0.47\textwidth]{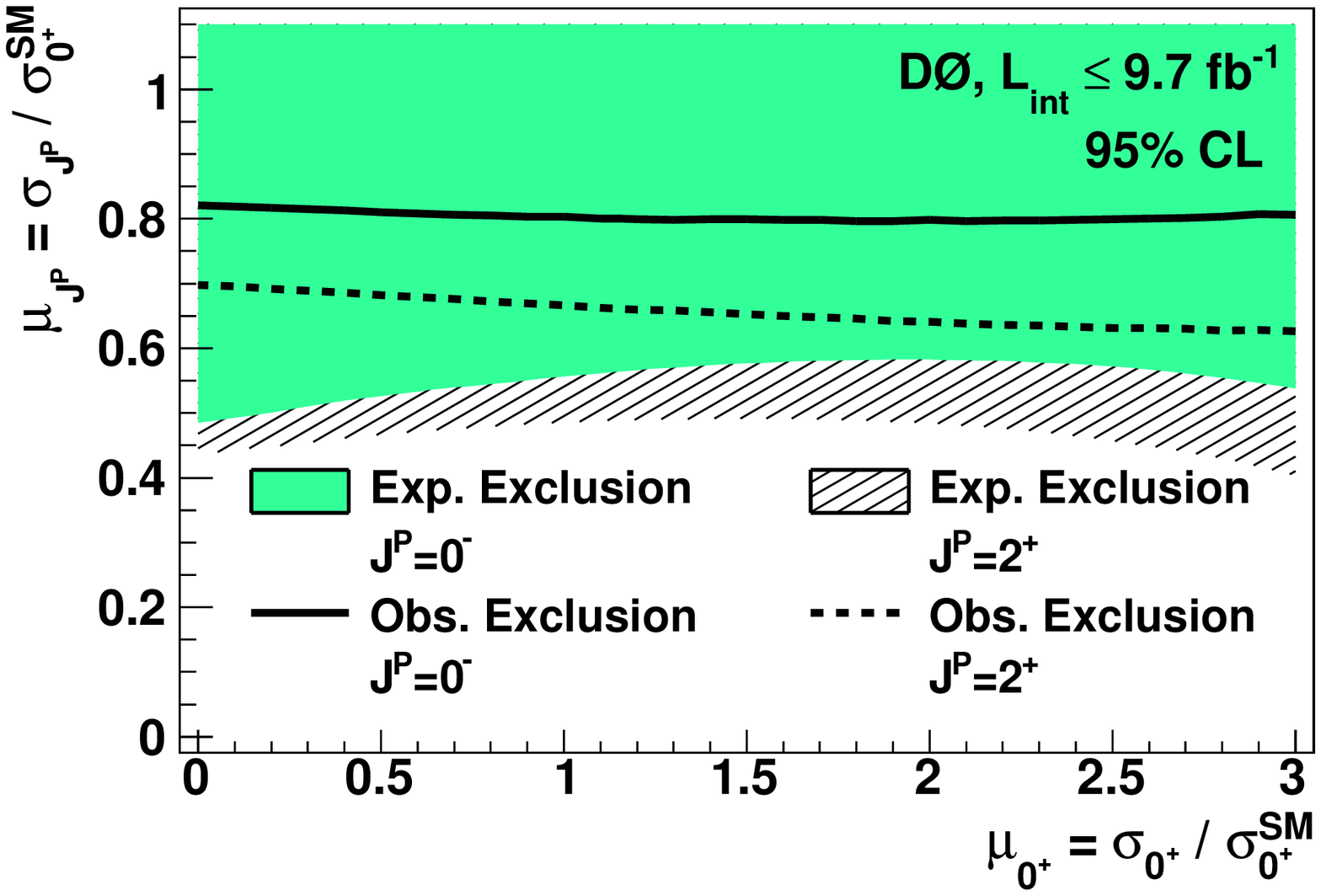}
\caption{(color online) The expected exclusion region (shaded area) 
and observed exclusion (solid line) as functions of the \jpzm\ and
\jpzp\ signal strengths. The expected exclusion region (hatched area)
and observed exclusion (dashed line) as functions of the
\jptp\ and \jpzp\ signal strengths.\label{twoDplot}}
\end{figure}

Tables detailing the $CL_{H_x}$ values for each individual analysis
channel and the combination can be found in Ref.~\cite{epaps}.  We
also obtain $1-CL_{s}$ over a range of SM and non-SM signal
strengths. Figure~\ref{twoDplot} shows the expected and observed 95\%\
CL exclusions as a function of the \jpzm\ (\jptp) and \jpzp\ signal
strengths, which may differ between the SM and non-SM signals. In the
tests shown in Fig.~\ref{twoDplot} the signal in the $H_1$ hypothesis
is the \jpzm (\jptp) signal normalized to
$\mu_{0^{-}}(\mu_{2^{+}})\times\sigma^{SM}_{0^{+}}$, and the signal in
the $H_0$ hypothesis is the \jpzp\ signal normalized to
$\mu_{0^{+}}\times\sigma^{SM}_{0^{+}}$.

\begin{figure}[t!]
\centering
\includegraphics[width=0.47\textwidth]{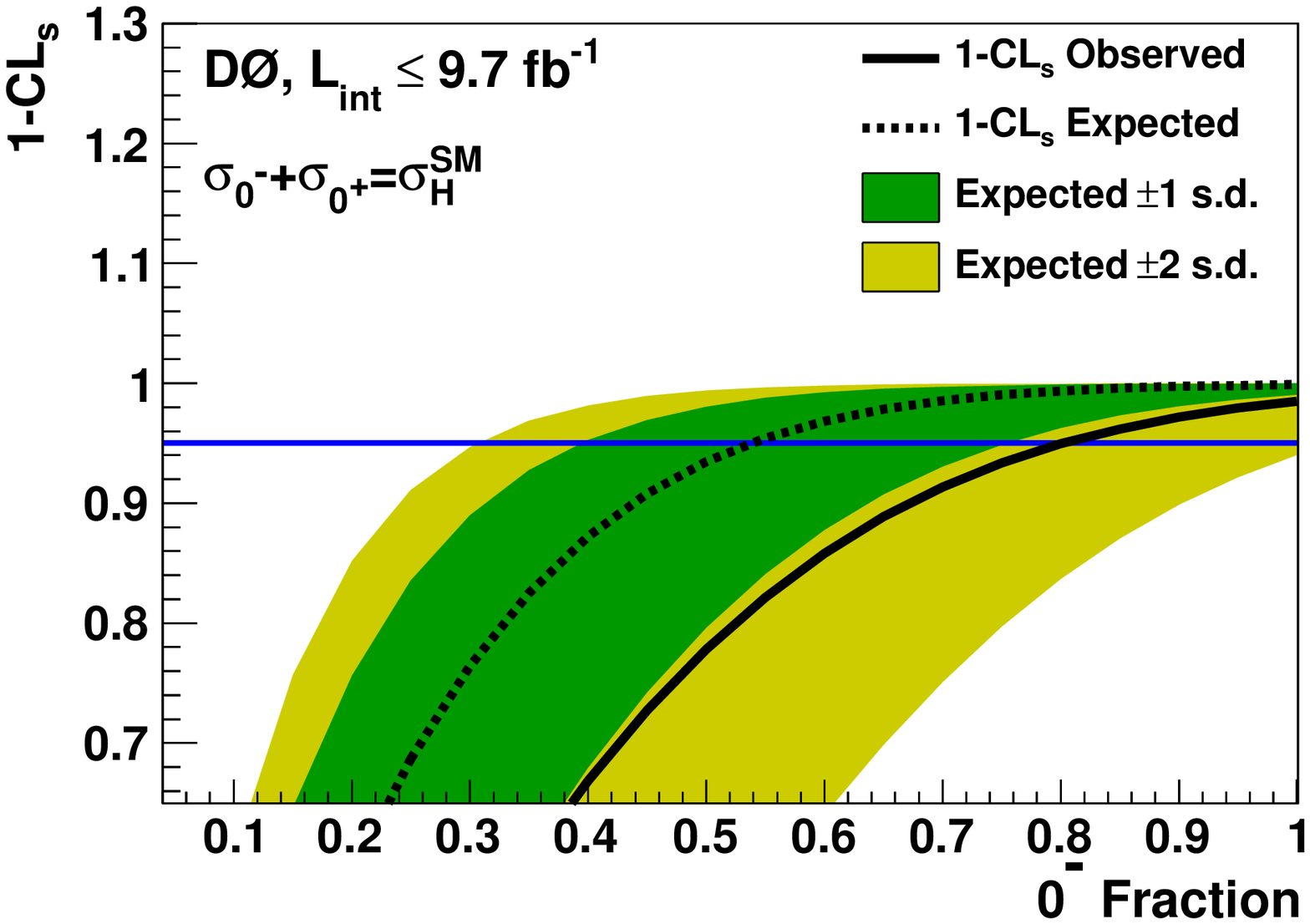}
\caption{\label{fig:sig_frac}(color online) $1-CL_{s}$ as a function 
of the \jpzm\ signal fraction assuming the product of the total cross
section and branching ratio is equal to the SM prediction. The
horizontal blue line corresponds to the 95\%\ CL exclusion. The dark
and light shaded regions represent the 1 and 2 standard deviations
(s.d.) fluctuations for the \jpzp\ hypothesis.}
\end{figure}
We also consider the possibility of a combination of \jp\ signals in
our data (e.g.,~\jpzp\ and \jpzm). These tests provide constraints on
a number of theoretical models such as those containing pseudoscalar
bosons in addition to a SM-like Higgs boson. For these studies we fix
the sum of the two cross sections to a specific value of $\mu \times
\sigma^{SM}_{0^{+}}$ and vary the fractions $f_{0^{-}} =
\sigma_{0^{-}}/(\sigma_{0^{+}}+\sigma_{0^{-}})$ or $f_{2^{+}} =
\sigma_{2^{+}}/(\sigma_{0^{+}}+\sigma_{2^{+}})$ of non-SM signal and
calculate the same $CL_{s}$ values as above as a function of
$\fzminus$ or $\ftplus$. To study $\fzminus$, we now modify $H_1$ to
be the sum of the background, the \jpzm\ signal normalized to
$\mu\times\sigma^{SM}_{0^{+}}\times\fzminus$, and the \jpzp\ signal
normalized to
$\mu\times\sigma^{SM}_{0^{+}}\times(1-\fzminus)$. $H_{0}$ remains as
previously defined. We follow an identical prescription for
\jptp. Figure~\ref{fig:sig_frac} presents the value $1-CL_{s}$ as a
function of the \jpzm\ signal fraction \fzminusm\ for the case of
$\mu=1.0$, and the corresponding figure for the \jptp\ hypothesis is
available in Ref.~\cite{epaps}. For $\mu=1.0$ we exclude a \jpzm\
(\jptp) signal fraction $\fzminus > 0.80$ ($\ftplus > 0.67$) at the
95\%\ CL. The expected exclusions are $\fzminus > 0.54$ ($\ftplus >
0.47$). Limits on admixture fractions for other choices of $\mu$ are 
shown in~\cite{epaps}.

In summary, we have performed tests of models with non-SM spin and parity
assignments in Higgs boson production with a $W$ or
$Z$ boson and decaying into $b\bar{b}$ pairs. We use the published
analyses of the \whl ,
\zhl, and \zhv\ final states with no modifications to the event selections. 
Sensitivity to non-SM \jp\ assignments in the two models considered
here is enhanced via the separation of samples into high- and
low-purity categories wherein the total mass or total transverse mass
of the$\,$\textit{VX}~system provides powerful discrimination.
Assuming a production rate compatible with both Tevatron and LHC Higgs
boson measurements, our data strongly reject non-SM \jp\ predictions,
and agree with the SM
\jpzp\ prediction. Under the assumption of two nearly degenerate bosons with different 
\jp\ values, we set upper limits on the fraction of non-SM signal in 
our data. This is the first exclusion of non-SM \jp\ parameter space
in a fermionic decay channel of the Higgs boson.
\begin{table}[htp]
\begin{center}
\begin{tabular}{c|@{\hskip 0.5cm}c@{\hskip 0.5cm}c@{\hskip 0.25cm}|@{\hskip 0.25cm}r@{\hskip 0.5cm}r}
\hline
\hline 
\jp\  &  \multicolumn{2}{c}{$1-CL_{s}$ (s.d.)} &  \multicolumn{2}{c}{$f_{J^{P}}$} \T\B \\ \hline
$\mu=1.0$  & Exp. & Obs. & Exp. & Obs. \T\B \\
\hline
$0^{-}$  &  0.9986 (3.00) &  0.976 (1.98) &  $>$0.54  &  $>$0.80 \T \\
$2^{+}$  &  0.9994 (3.22) &  0.990 (2.34) &  $>$0.47  &  $>$0.67 \B \\ \hline
$\mu=1.23$ &  &  &  & \T\B \\ \hline
$0^{-}$  &  0.9998 (3.60) &  0.995 (2.56) &  $>$0.45  &  $>$0.67 \T \\
$2^{+}$  &  0.9999 (3.86) &  0.998 (2.91) &  $>$0.40  &  $>$0.56 \B \\
\hline
\hline
\end{tabular}
\caption{Expected and observed $1-CL_{s}$ values (converted to s.d.\ in parentheses) and signal fractions for $\mu=1.0$ 
and $\mu=1.23$ excluded at the 95\%\ CL.\label{tab:sum}}
\end{center}
\end{table}

%
We thank the staffs at Fermilab and collaborating institutions,
and acknowledge support from the
DOE and NSF (USA);
CEA and CNRS/IN2P3 (France);
MON, NRC KI and RFBR (Russia);
CNPq, FAPERJ, FAPESP and FUNDUNESP (Brazil);
DAE and DST (India);
Colciencias (Colombia);
CONACyT (Mexico);
NRF (Korea);
FOM (The Netherlands);
STFC and the Royal Society (United Kingdom);
MSMT and GACR (Czech Republic);
BMBF and DFG (Germany);
SFI (Ireland);
The Swedish Research Council (Sweden);
and
CAS and CNSF (China).

\bibliography{higgs}
\bibliographystyle{h-physrev3}

\begin{widetext}
\section*{SUPPLEMENTAL MATERIAL}

In this document we provide supplemental information on the 
constraints on models with non-SM spin and parity for the Higgs 
boson in the \vbb\ final states in up to $9.7$ \infb\ of 
$p\bar{p}$ collisions at $\sqrt{s} = $ 1.96~TeV collected 
with the D0 detector at the Fermilab Tevatron Collider. We denote a non-SM 
Higgs boson as $X$.

\appendix

\begin{description}
\item[Figure~\ref{fig:dijetm}:] Dijet mass distributions for the \vvbb\ and \llbb\ analyses and the BDT output distribution for the \lvbb\ analysis.
\item[Figures~\ref{fig:llbb_inout}--\ref{fig:nunubb_dataMCinout}:] Additional$\,$\textit{VX} invariant and transverse mass distributions 
for individual analyses.
\item[Figures~\ref{fig:llr_zm} and~\ref{fig:llr_tp}:] LLR distributions for the individual analyses and 
their combination.
\item[Tables~\ref{table:cls} and~\ref{table:cls_1.23}:] Tables of $CL_{H_{x}}$ and $1-CL_{s}$ values for the individual analyses 
and their combination for $\mu=1.0$ and $\mu=1.23$.
\item[Figure~\ref{fig:sig_frac_jptp}:] $1-CL_{s}$ as a function of the \jptp\ signal fraction, 
\ftplusm, for all analyses combined.
\item[Figure~\ref{fig:fracrate}:] The expected and observed 95\%\ CL exclusion as functions of the \jpzm\ (\jptp) signal fraction, \fzminusm\ (\ftplusm), 
and the total signal strength.
\end{description}

\begin{figure}[b!]
\centering
\includegraphics[width=0.47\textwidth]{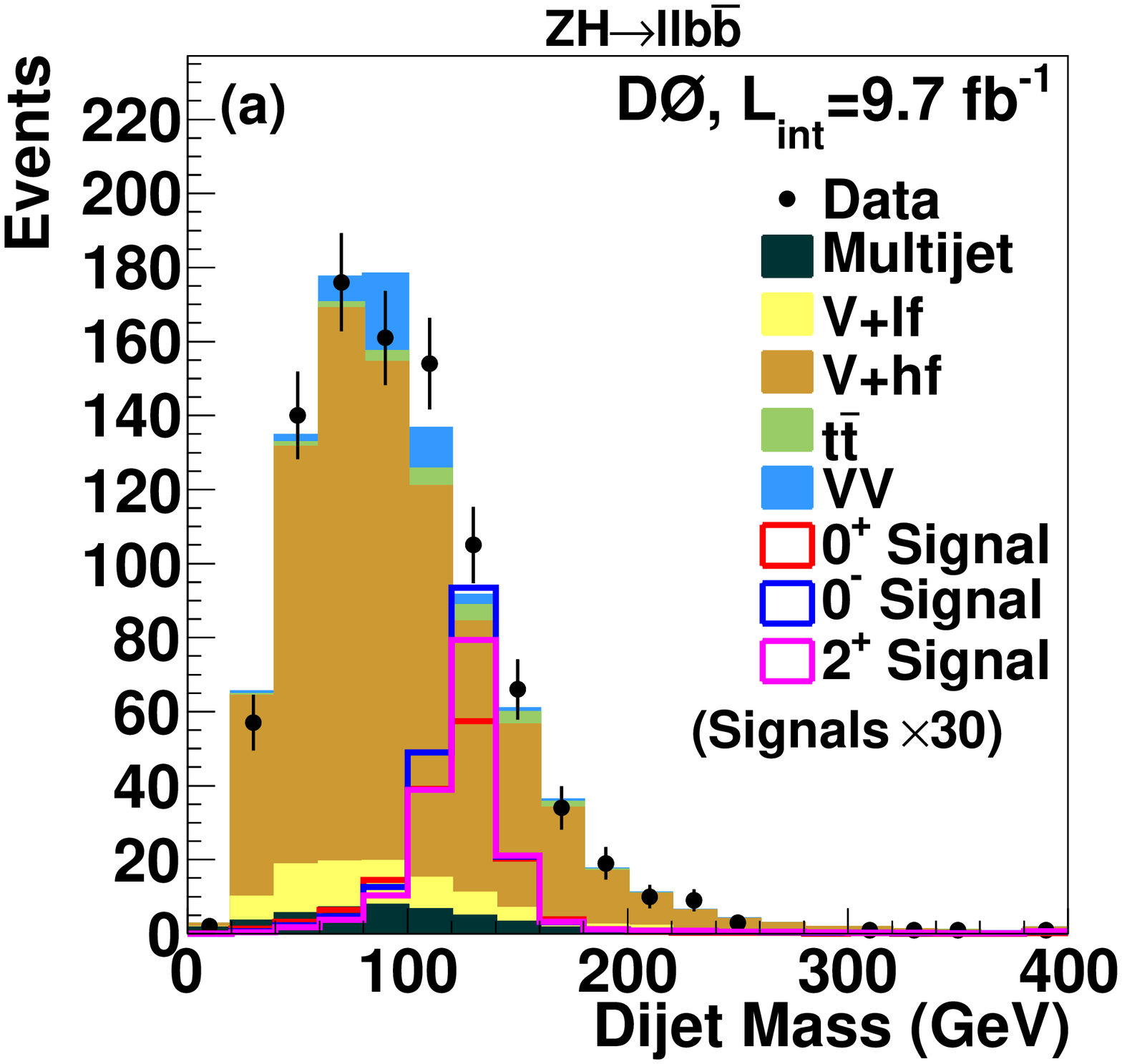}
\hspace{5 mm}
\includegraphics[width=0.47\textwidth]{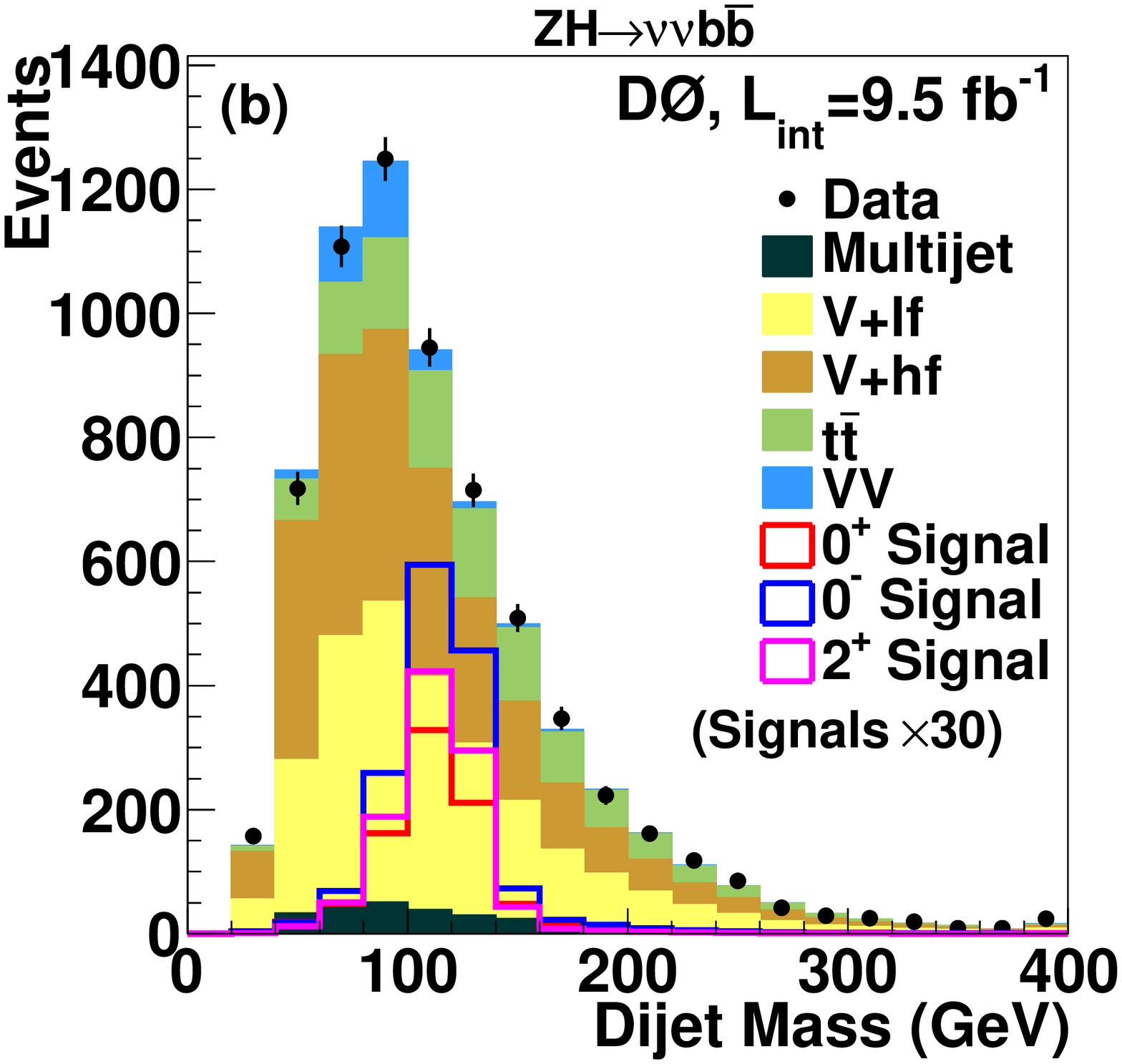}

\vspace{8 mm}

\includegraphics[width=0.47\textwidth]{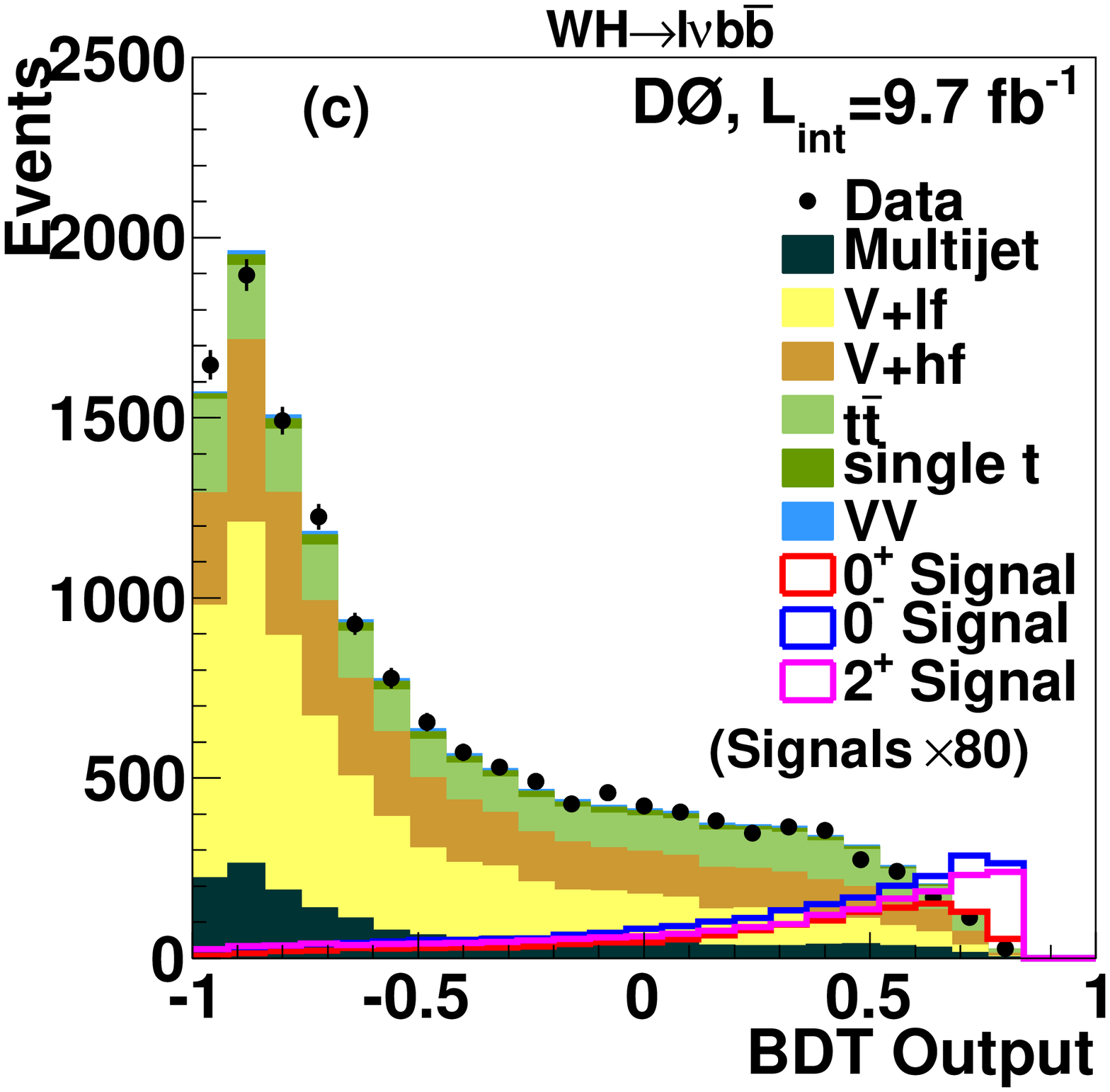}
\caption{Invariant mass of the dijet system for (a) the \zhl\ analysis, and (b) the \zhv\ analysis, and the BDT output for (c) the \whl\ analysis. 
The \jptp\ and \jpzm\ 
samples are normalized to the product of the SM cross section and branching fraction multiplied by an additional factor. Heavy- and light-flavor quark jets are denoted by 
lf and hf, respectively. Overflow events are included in the highest bin.
For all signals, a mass of 125~GeV for the~$\!H$ or~$\!X$ boson is assumed.
\label{fig:dijetm}}
\end{figure}

\begin{figure}[b!]
\centering
\includegraphics[width=0.47\textwidth]{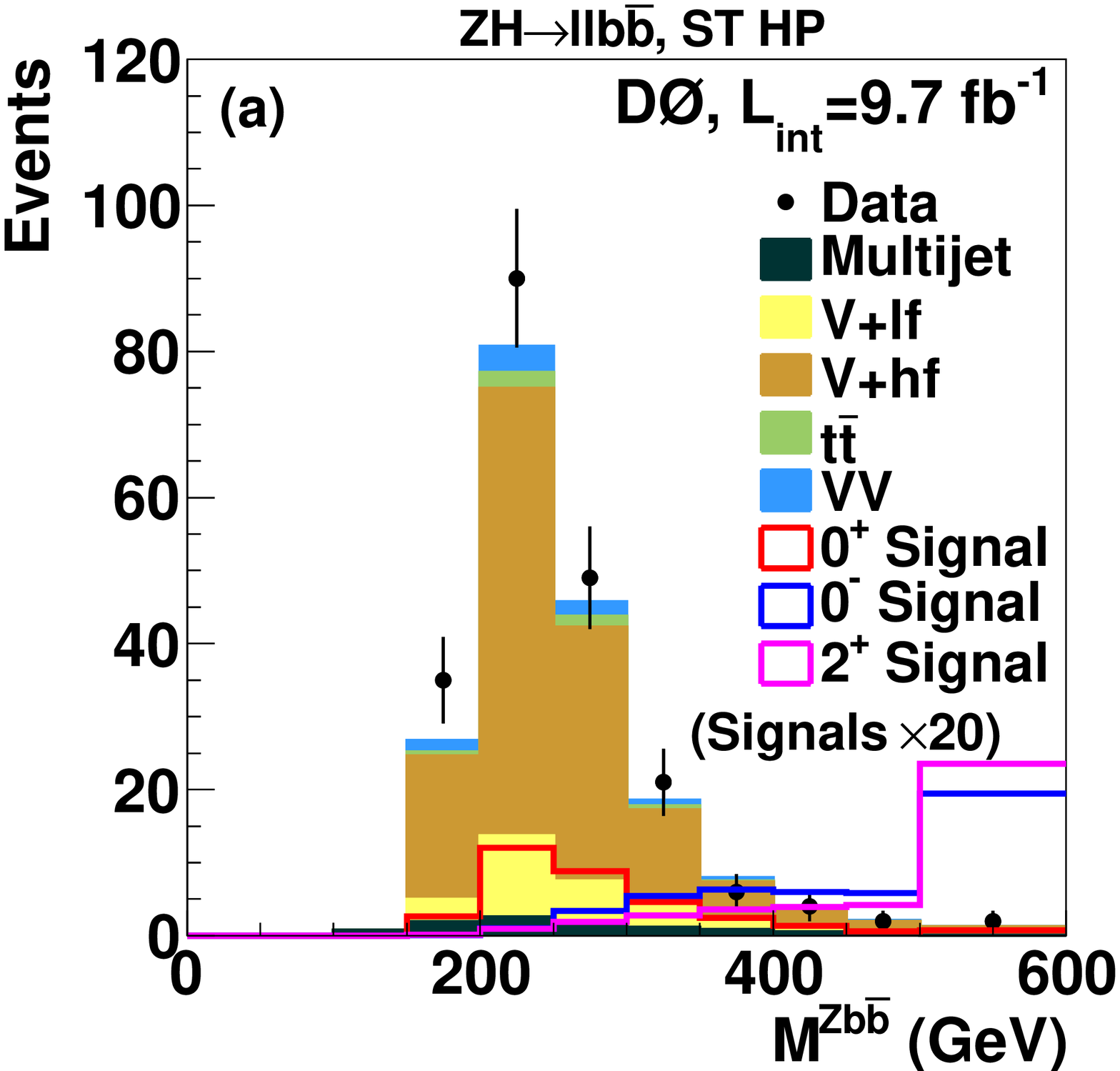}
\hspace{5 mm}
\includegraphics[width=0.47\textwidth]{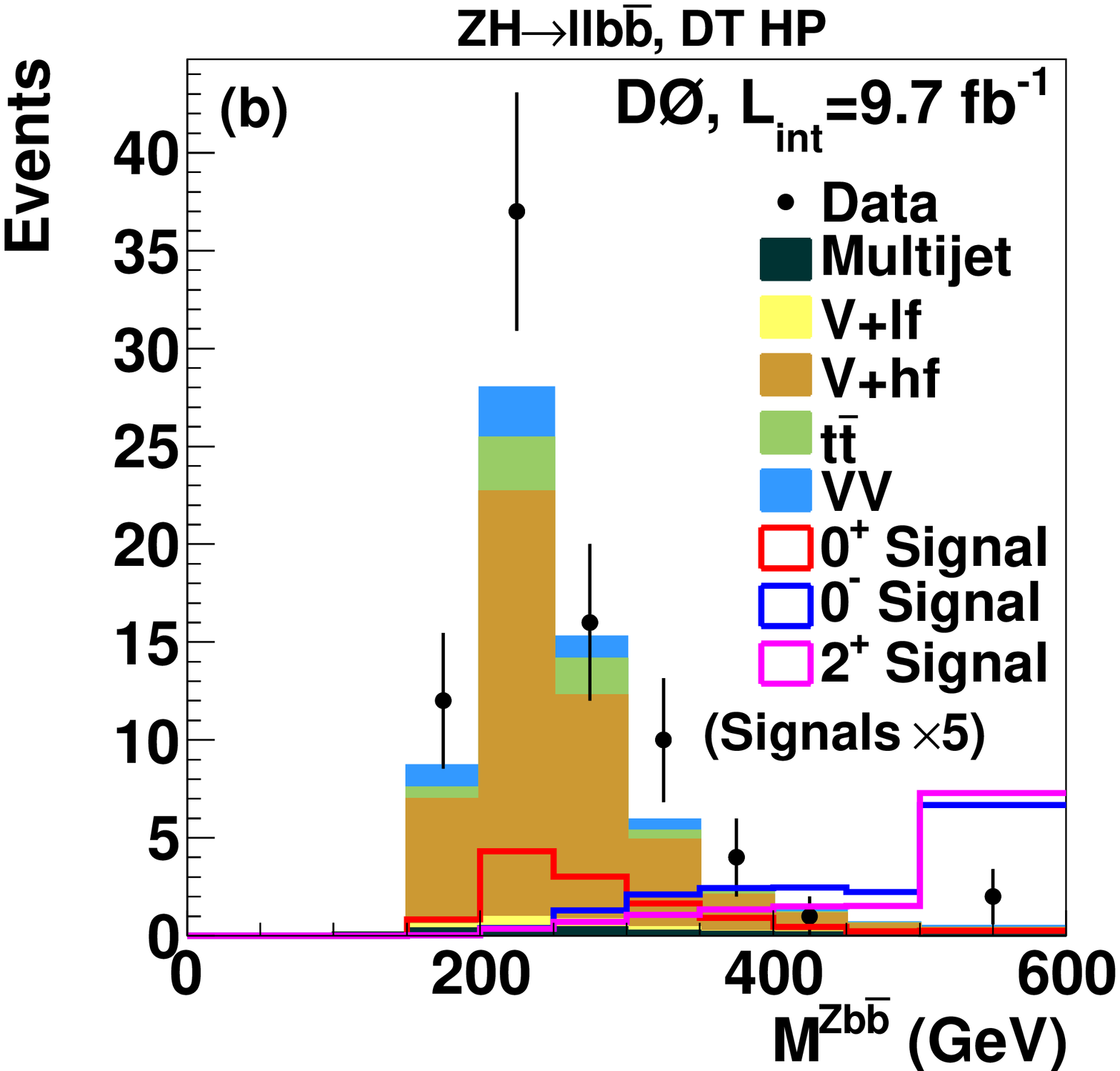}

\vspace{8 mm}

\includegraphics[width=0.47\textwidth]{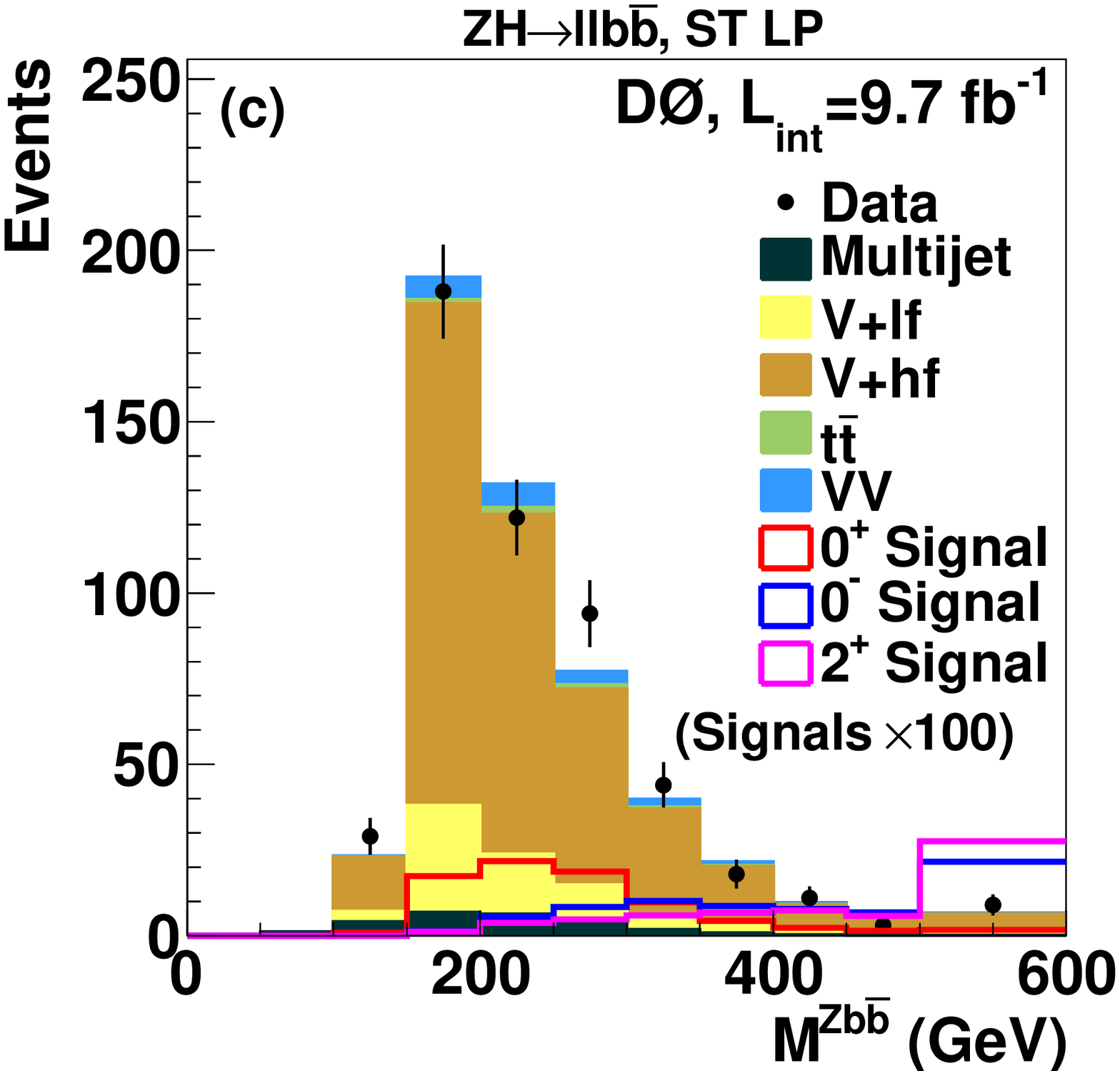}
\hspace{5 mm}
\includegraphics[width=0.47\textwidth]{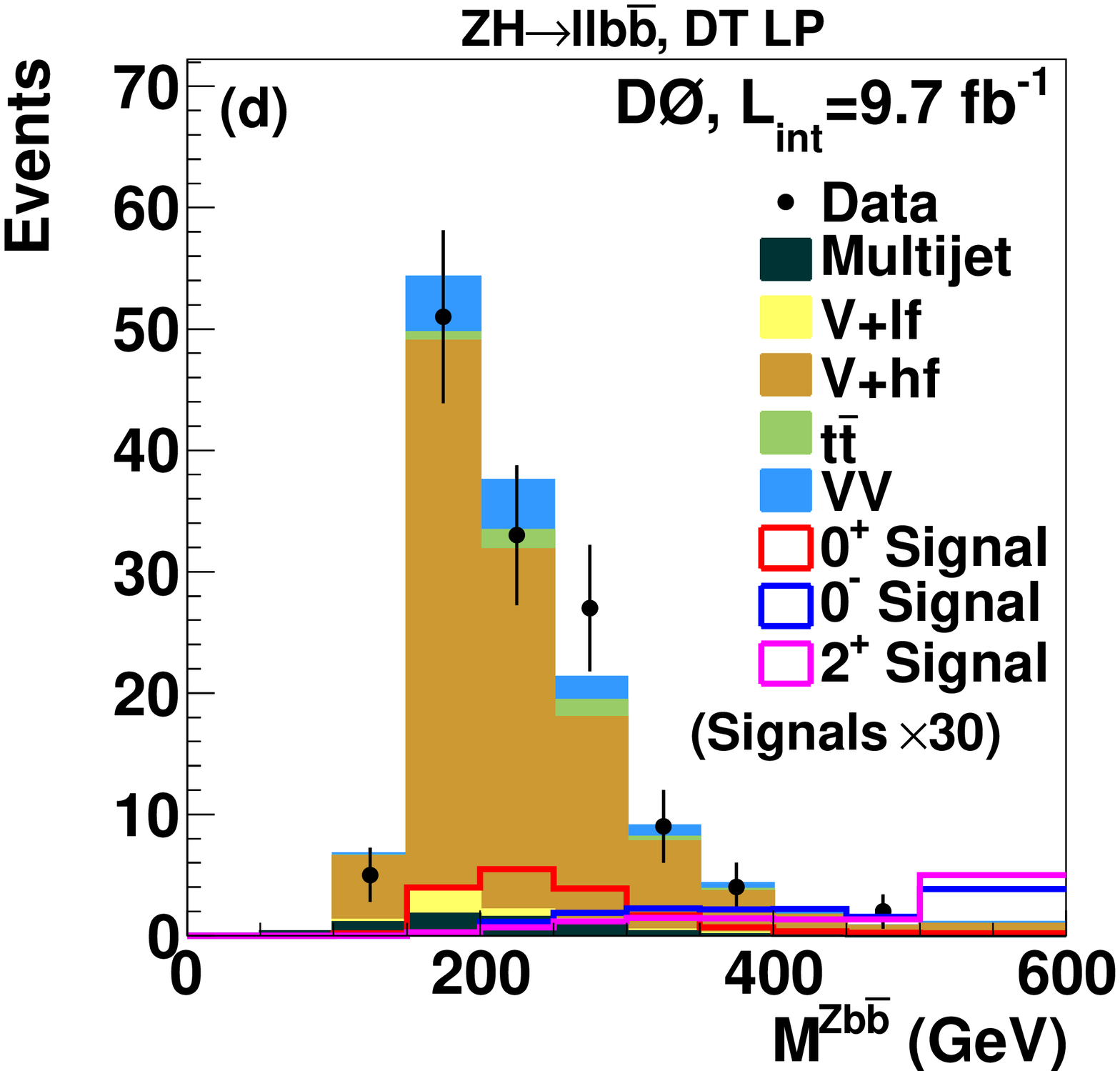}
\caption{Invariant mass of the $\ell\ell b\bar{b}$ system in the \zhl\ analysis for events in the 
(a) single-tag high-purity (ST HP), (b) double-tag  high-purity (DT HP), 
(c) single-tag low-purity (ST LP), and (d) double-tag low-purity (DT LP) channels. The \jptp\ and \jpzm\ 
samples are normalized to the product of the SM cross section and branching fraction multiplied by an additional factor. Heavy- and light-flavor 
quark jets are denoted by lf and hf, respectively. Overflow events are included in the last bin.
For all signals, a mass of 125~GeV for the~$\!H$ or~$\!X$ boson is assumed.
\label{fig:llbb_inout}}
\end{figure}

\begin{figure}[htbp]
\centering
\includegraphics[width=0.47\textwidth]{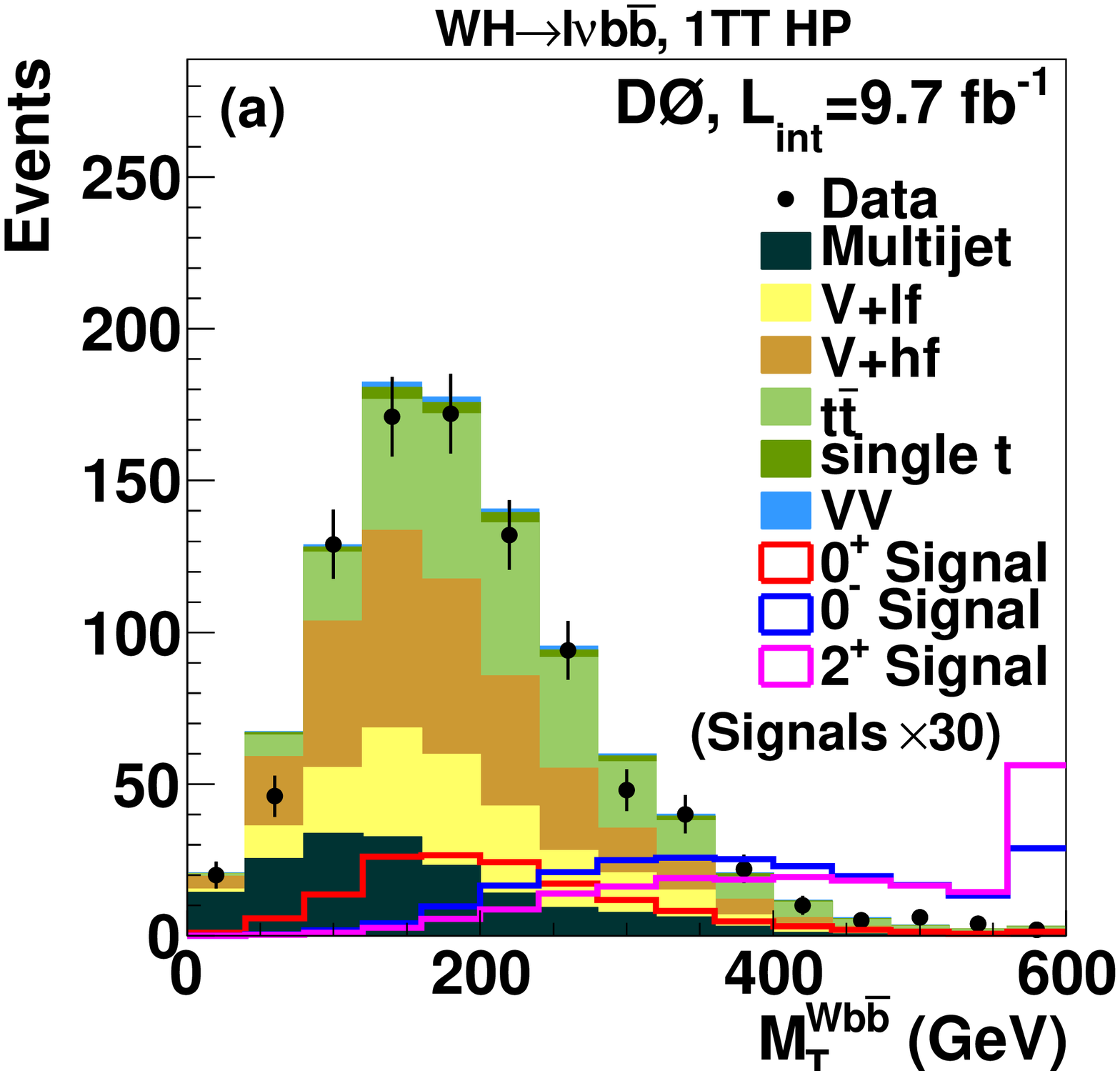}
\hspace{5 mm}
\includegraphics[width=0.47\textwidth]{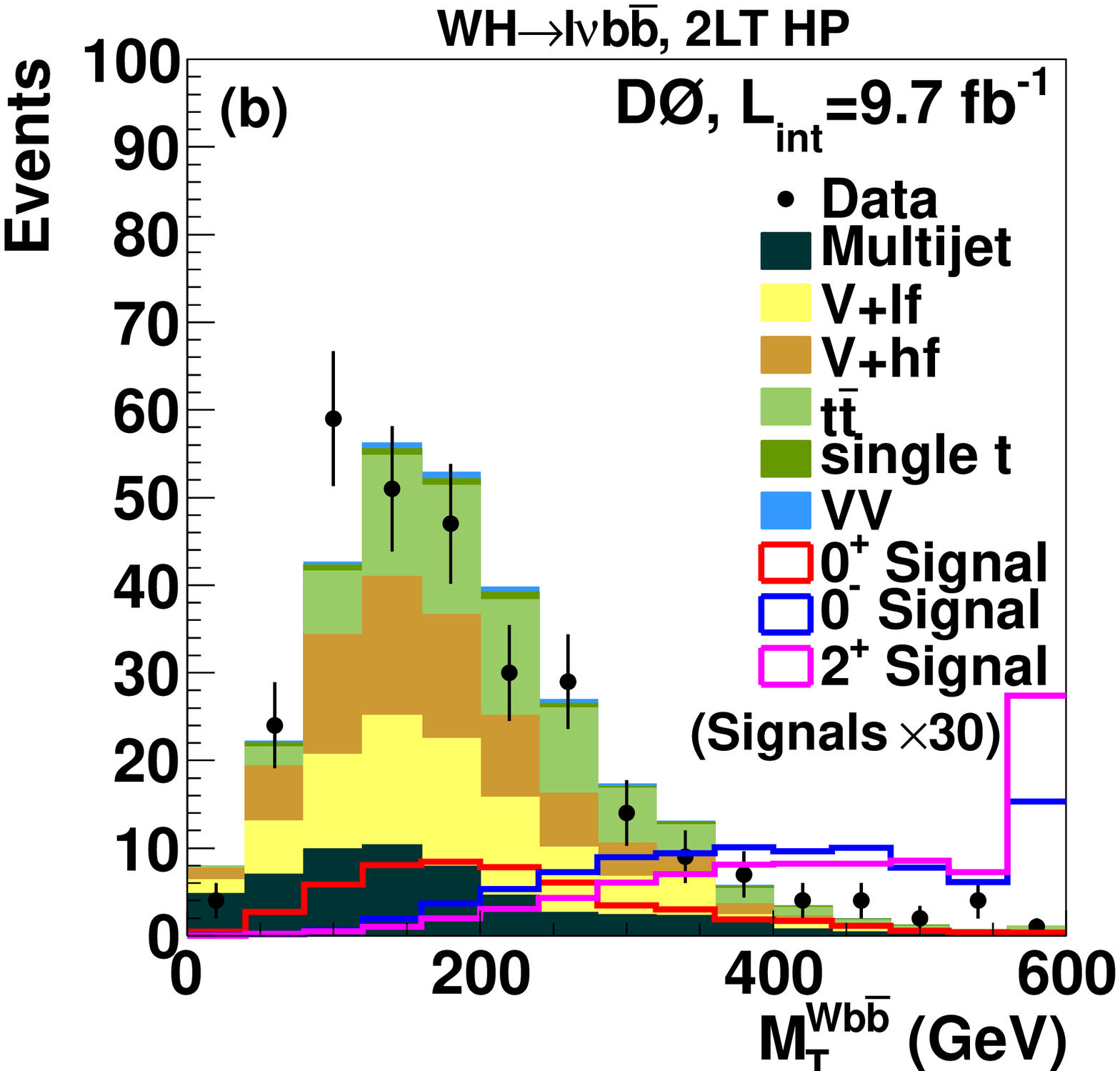}

\hspace{8 mm}

\includegraphics[width=0.47\textwidth]{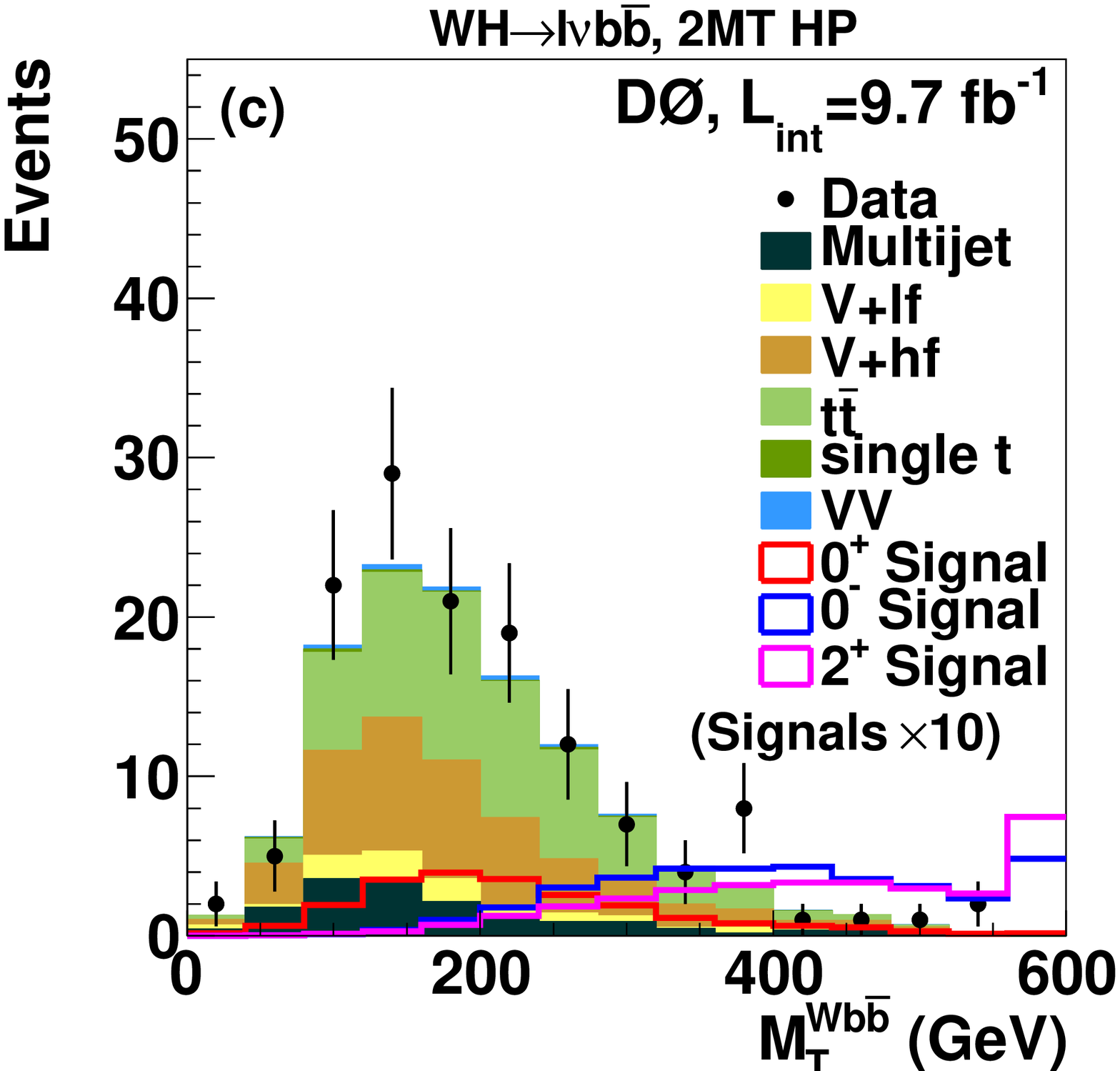}
\hspace{5 mm}
\includegraphics[width=0.47\textwidth]{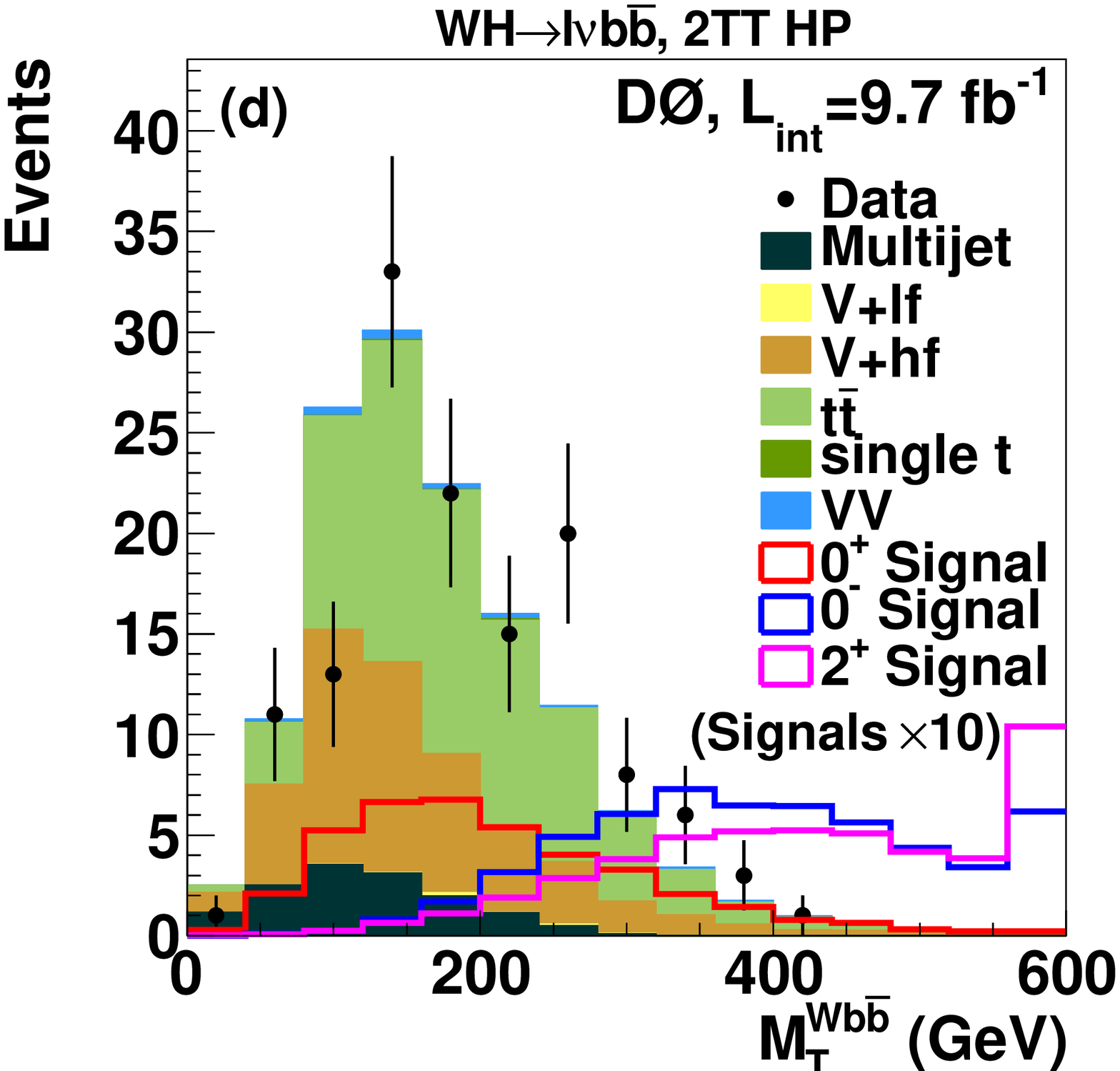}
\caption{Transverse mass of the $\ell\nu b\bar{b}$ system in the \whl\ analysis in the 
high-purity (HP) region for (a) 1 tight-tag (1TT), (b) 2 loose-tags (2LT), (c) 2 medium-tags (2MT), 
and (d) 2 tight-tags (2TT) channels.
 The \jptp\ and \jpzm\ samples are normalized to the product of the SM cross section and branching fraction multiplied by an additional factor. 
 Heavy- and light-flavor quark jets are denoted by lf and hf, respectively. Overflow 
events are included in the last bin. For all signals, a mass of 125~GeV for the~$\!H$ or~$\!X$ boson is assumed.
\label{fig:whl_dataMCinout_5}}
\end{figure}

\begin{figure}[htbp]
\centering
\includegraphics[width=0.47\textwidth]{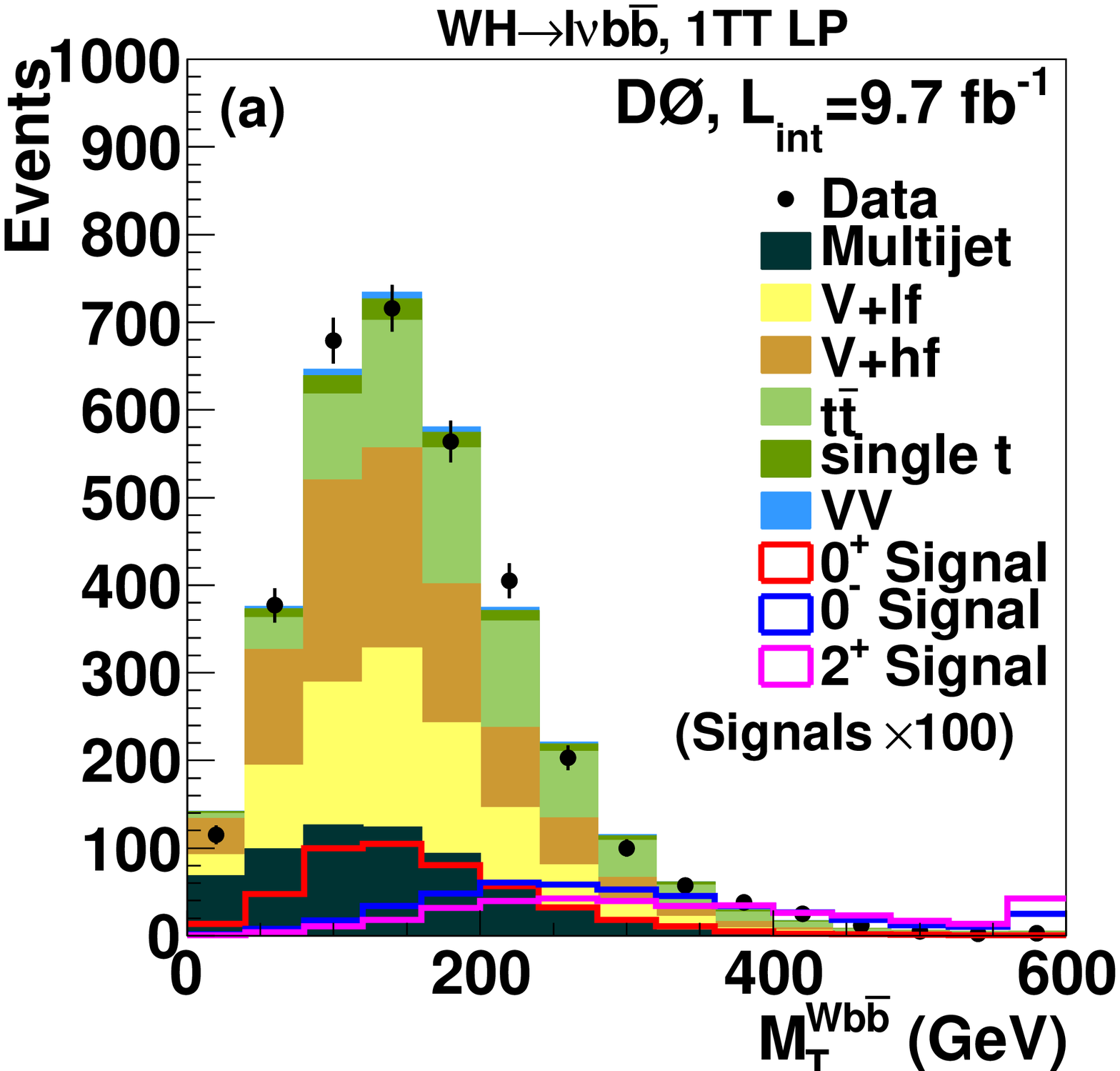}
\hspace{5 mm}
\includegraphics[width=0.47\textwidth]{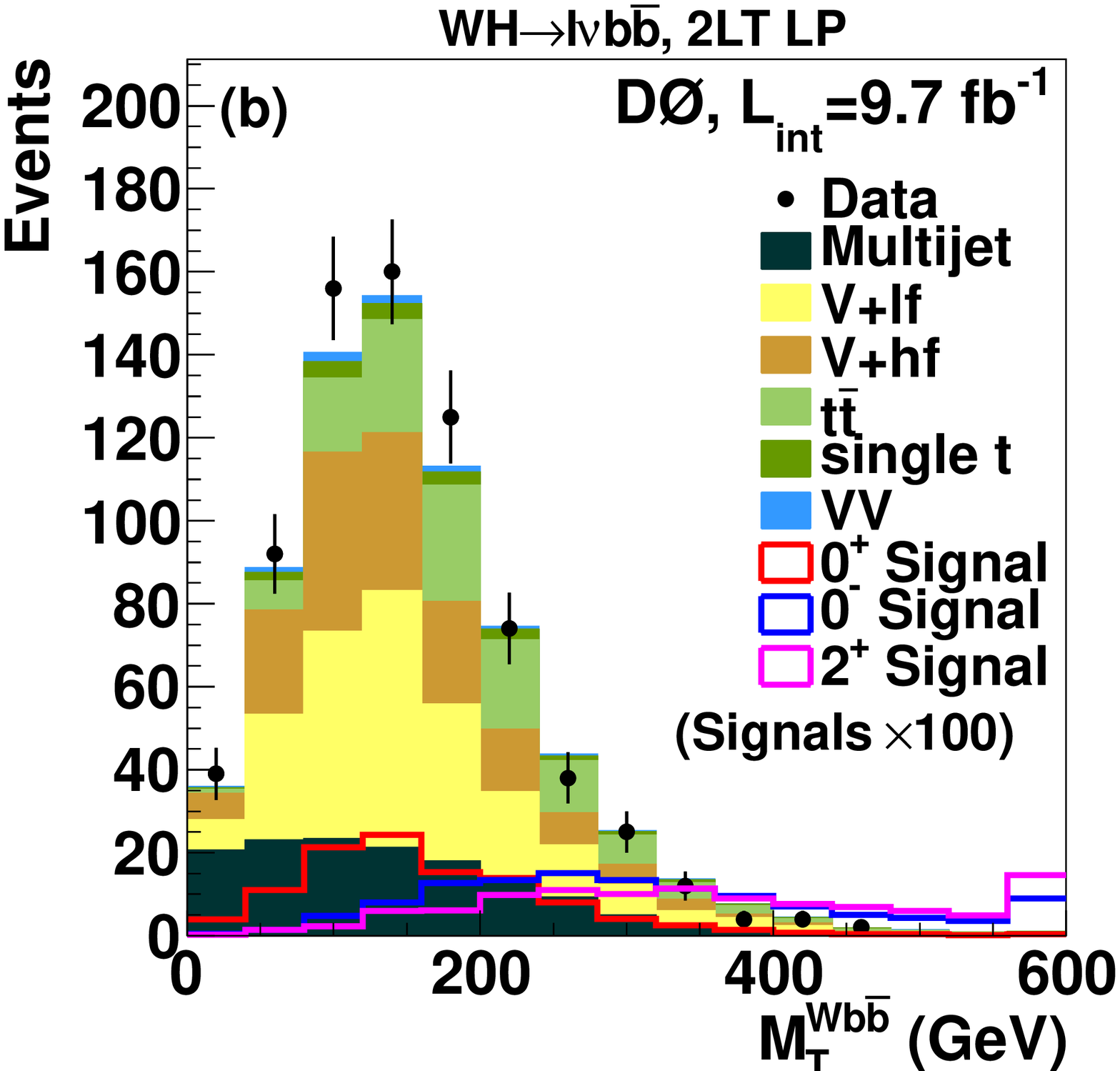}

\hspace{8 mm}

\includegraphics[width=0.47\textwidth]{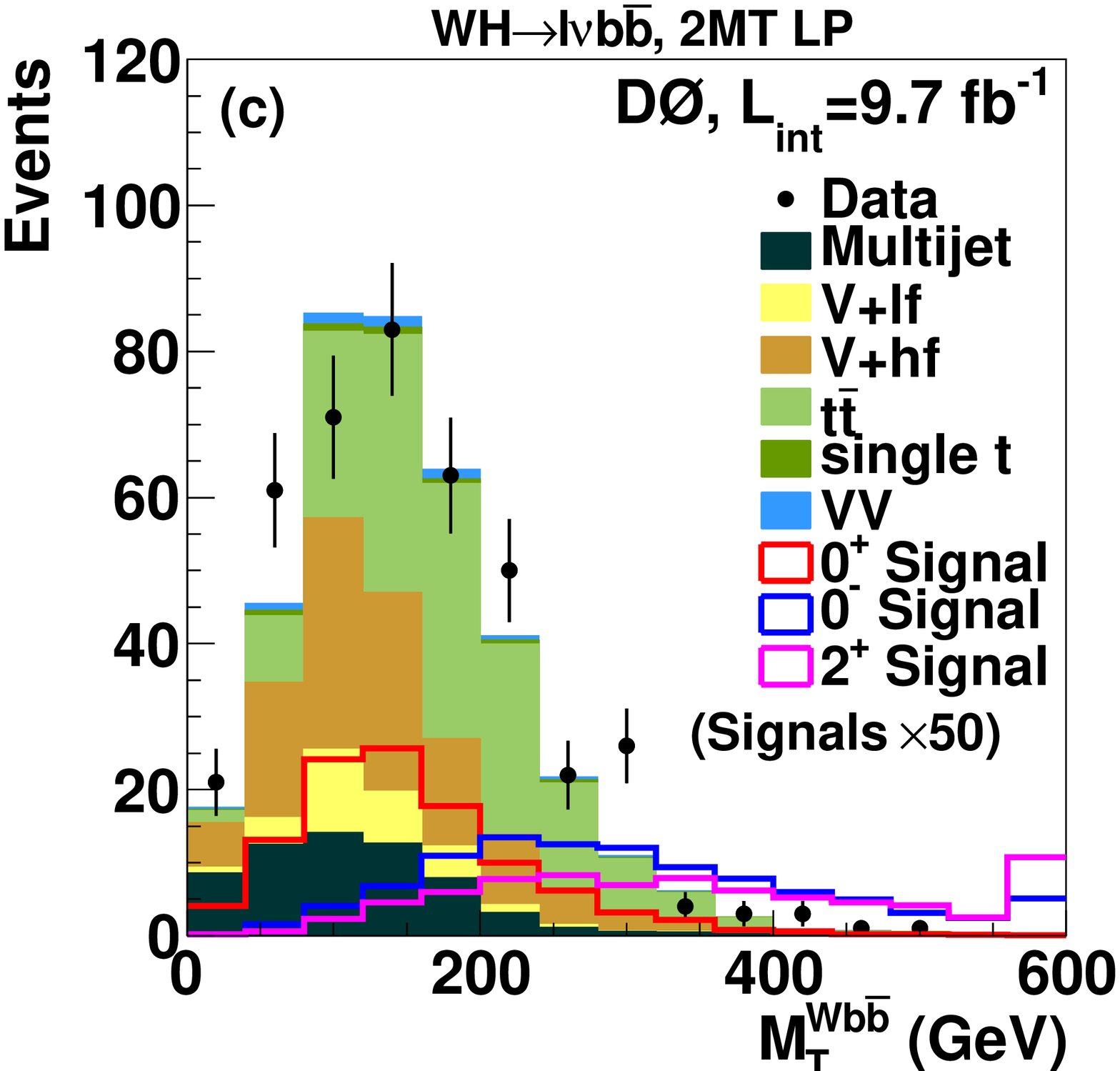}
\hspace{5 mm}
\includegraphics[width=0.47\textwidth]{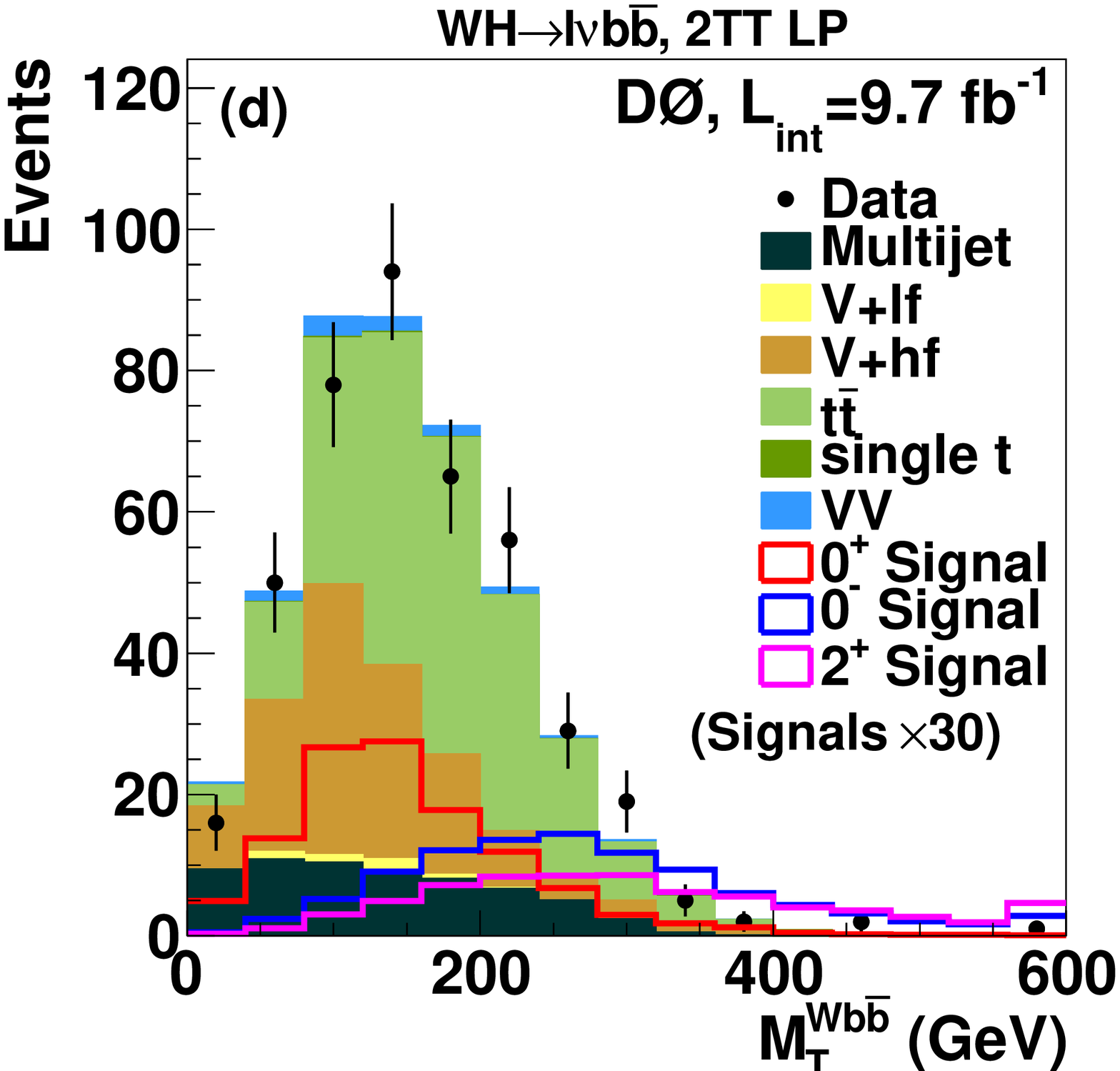}
\caption{Transverse mass of the $\ell\nu b\bar{b}$ system in the \whl\ analysis in
the low purity (LP) region for (a) 1-tight-tag (1TT), (b) 2-loose-tags (2LT), (c) 2-medium-tags (2MT),
and (d) 2-tight-tags (2TT) channels.
The \jptp\ and \jpzm\ samples are normalized to the product of the SM cross section and branching fraction multiplied by an additional factor. 
Heavy- and light-flavor quark jets are denoted by lf and hf, respectively. Overflow 
events are included in the last bin. For all signals, a mass of 125~GeV for the~$\!H$ or~$\!X$ boson is assumed.
\label{fig:whl_dataMCinout_0_5}}
\end{figure}

\begin{figure}[htbp]
\centering
\includegraphics[width=0.47\textwidth]{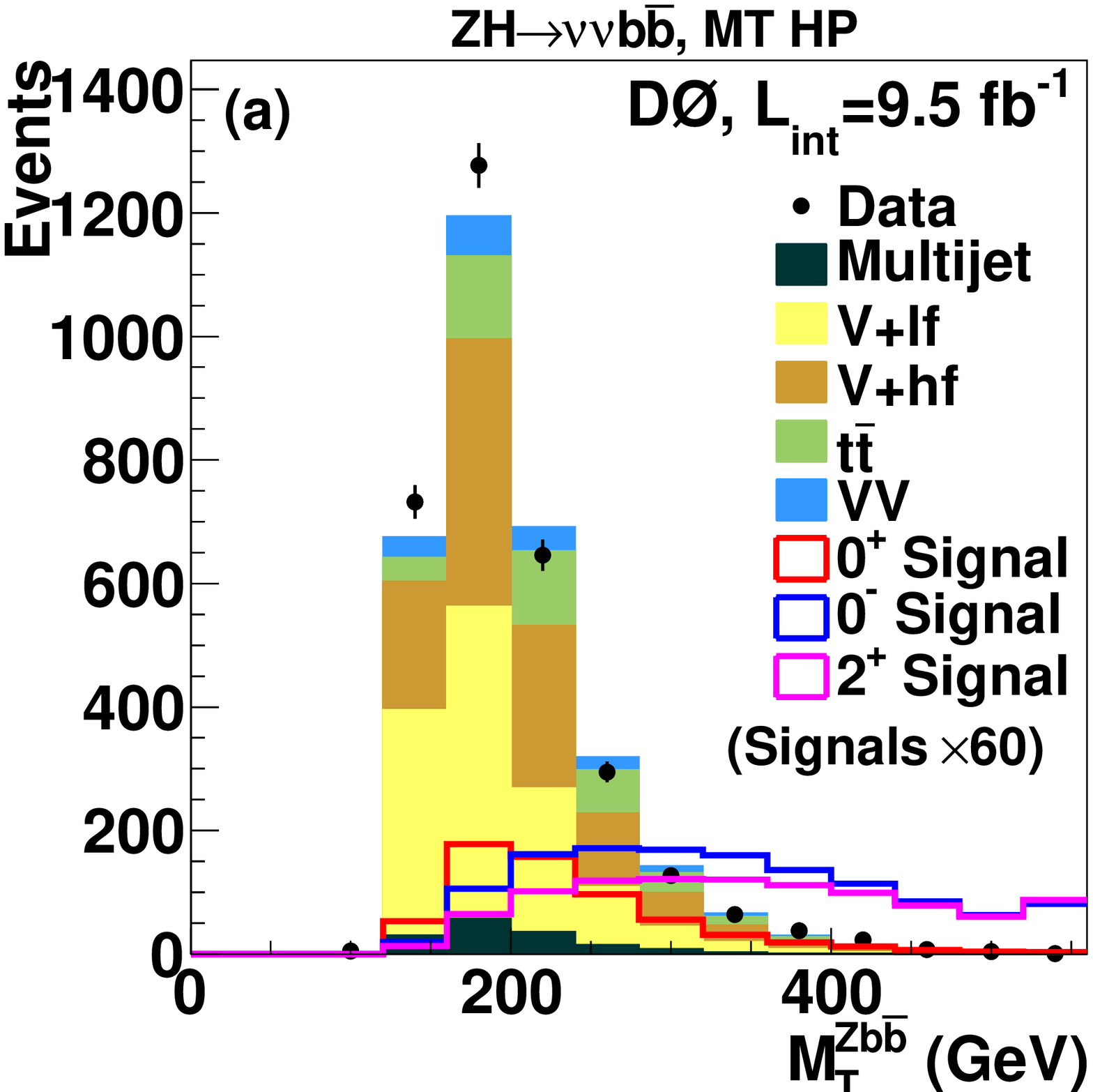}
\hspace{5 mm}
\includegraphics[width=0.47\textwidth]{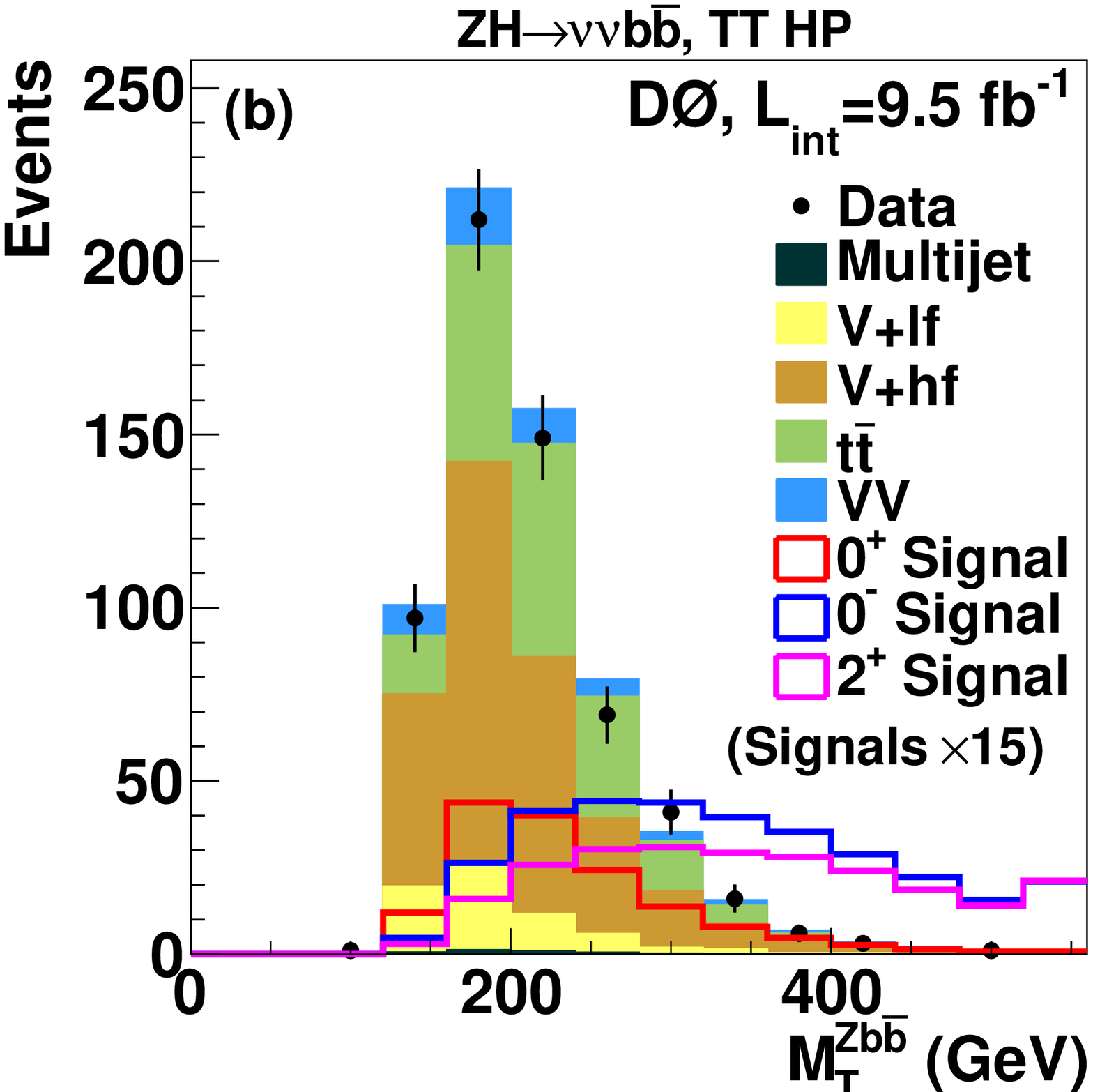}

\hspace{8 mm}

\includegraphics[width=0.47\textwidth]{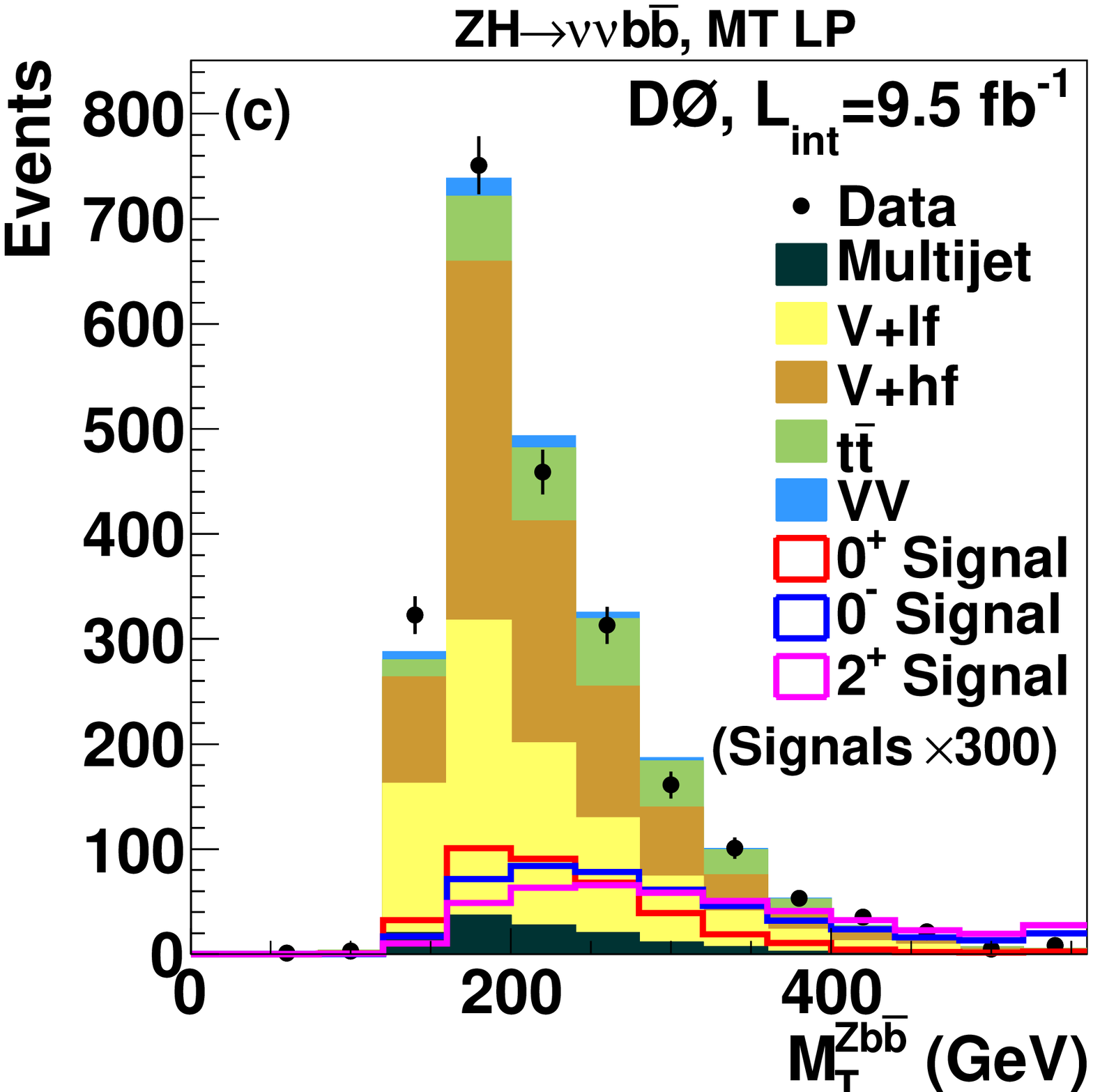}
\hspace{5 mm}
\includegraphics[width=0.47\textwidth]{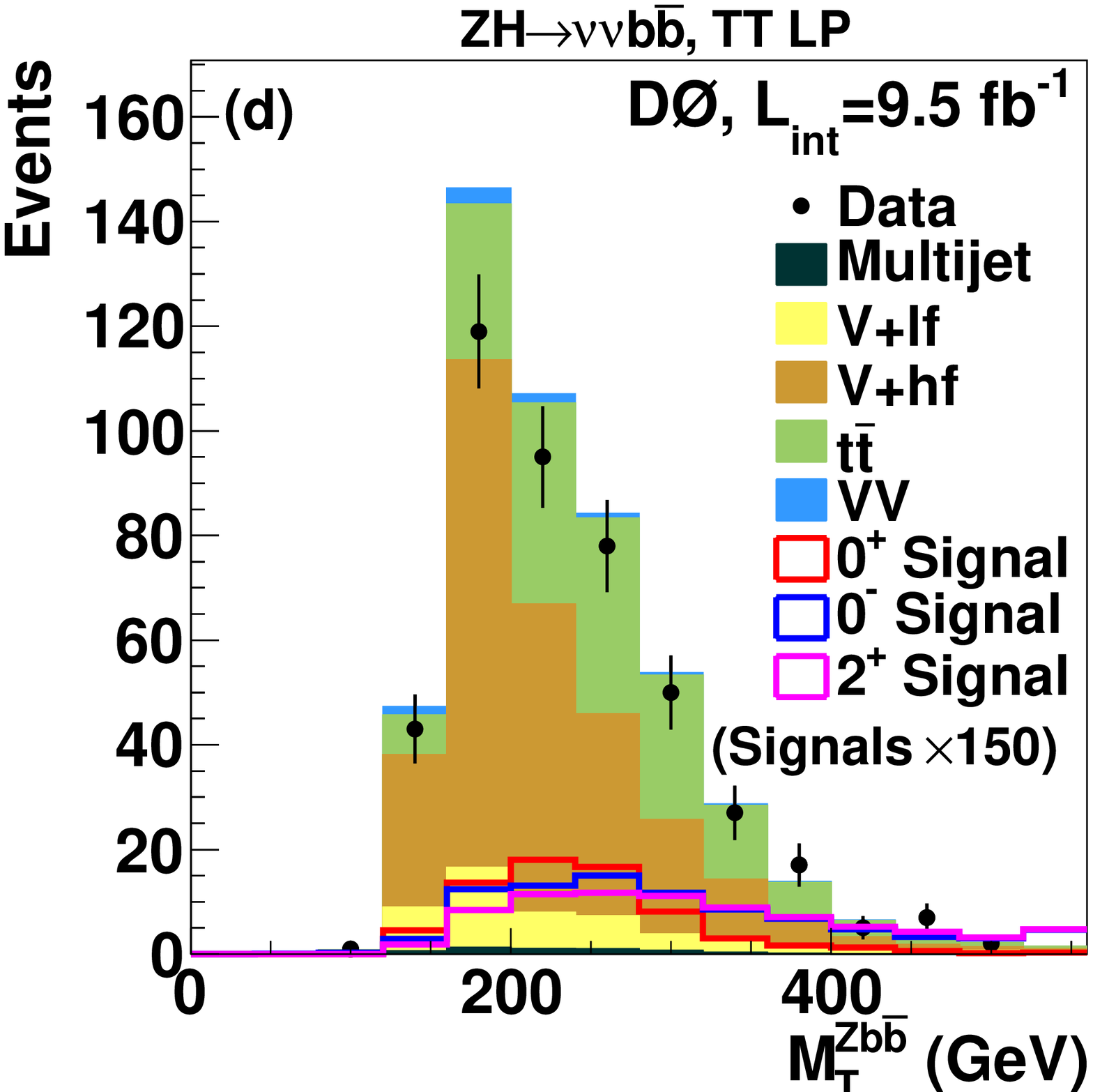}
\caption{Transverse mass of the $\nu\nu b\bar{b}$ system in the \zhv\ analysis for events in the (a) medium-tag 
high-purity (MT HP), 
(b) tight-tag high-purity (TT HP), (c) medium-tag low-purity (MT LP), and (d) tight-tag low-purity (TT LP) 
channels. The \jptp\ and \jpzm\ samples are normalized to the product of the SM cross section and branching fraction multiplied by an additional factor. 
Heavy- and light-flavor quark jets are denoted by lf and hf, respectively. Overflow 
events are included in the last bin. For all signals, a mass of 125~GeV for the~$\!H$ or~$\!X$ boson is assumed.
\label{fig:nunubb_dataMCinout}}
\end{figure}

\clearpage

\begin{figure}[htbp]
\centering
\includegraphics[width=0.47\textwidth]{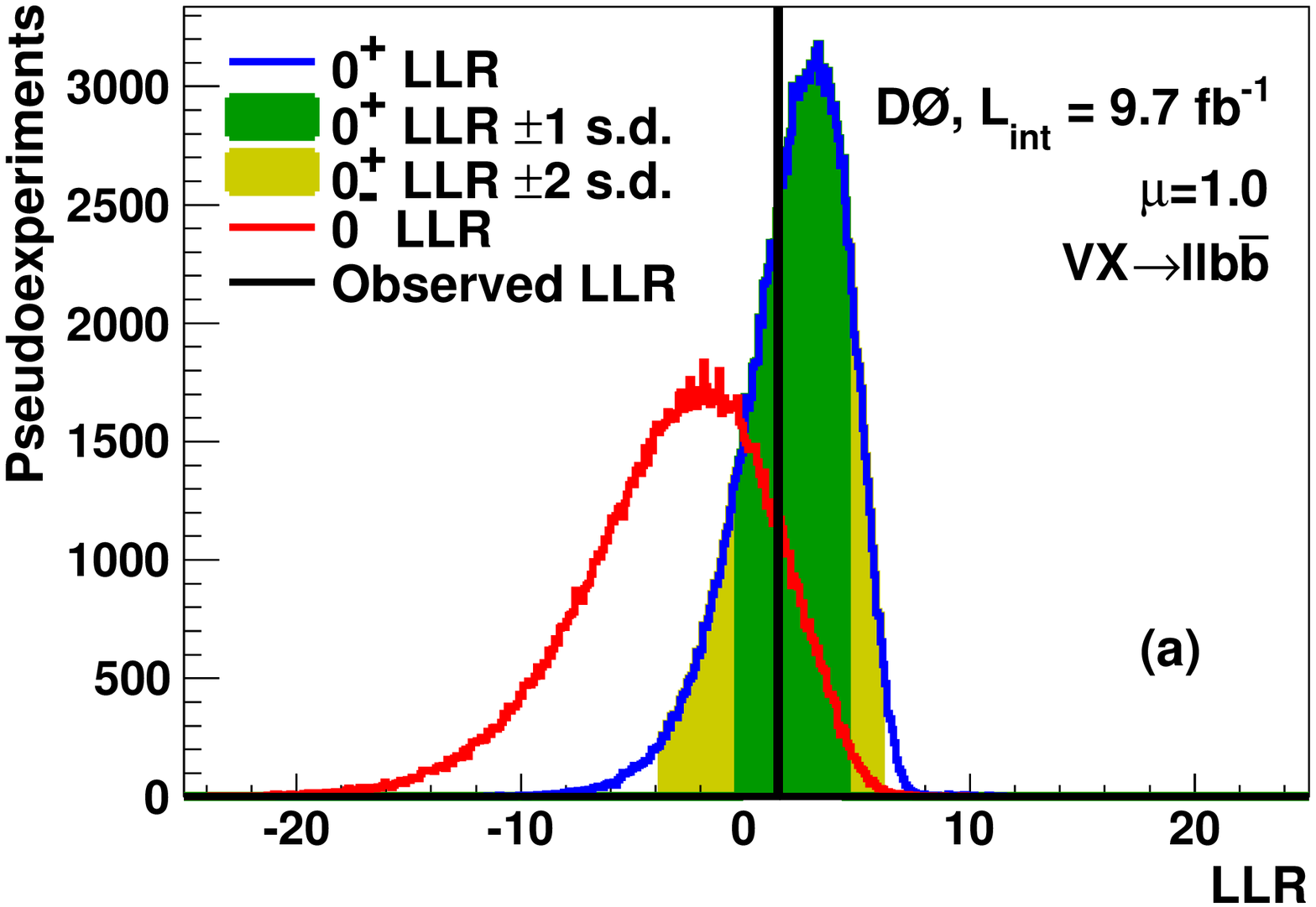}
\hspace{5 mm}
\includegraphics[width=0.47\textwidth]{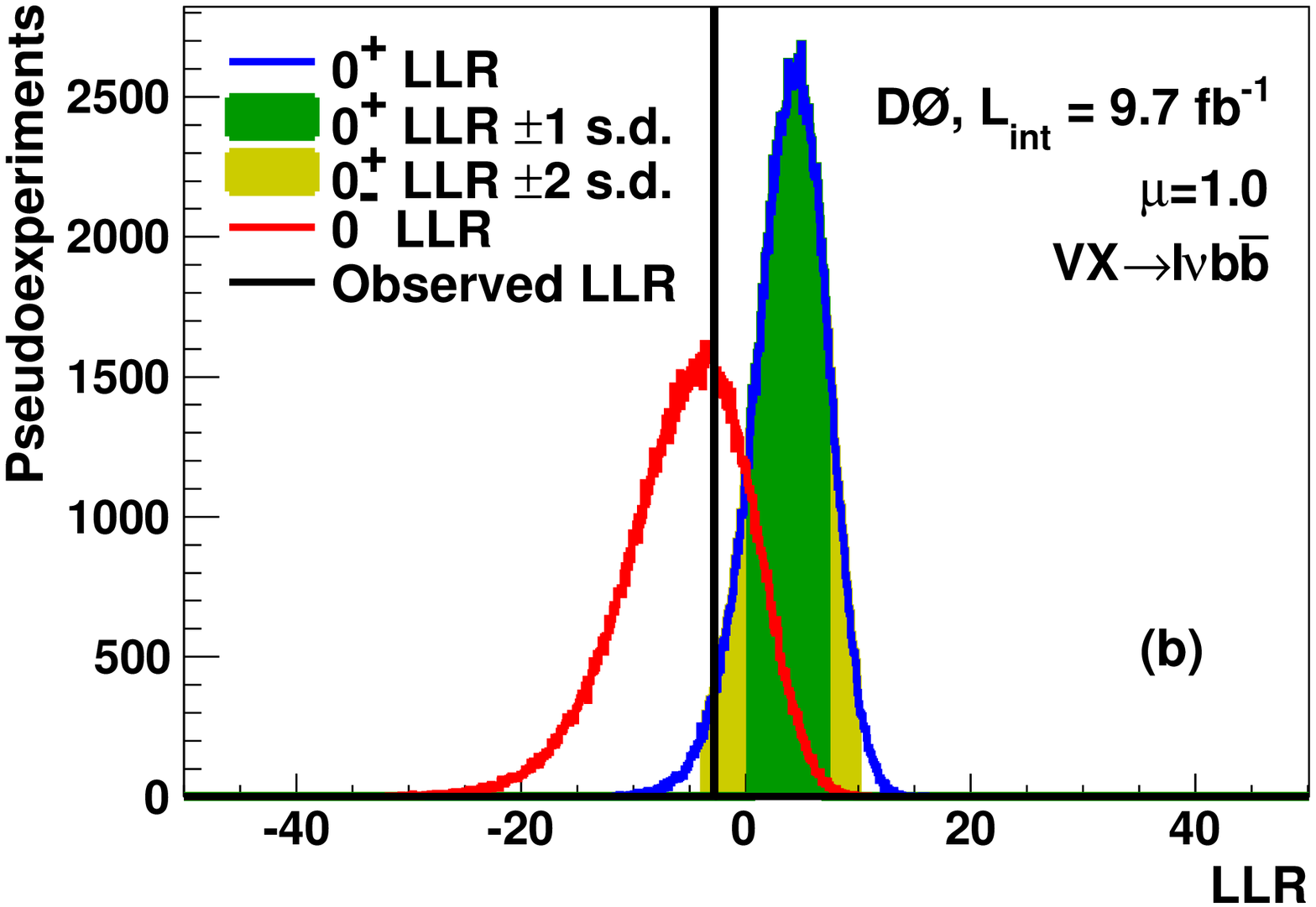}

\hspace{8 mm}

\includegraphics[width=0.47\textwidth]{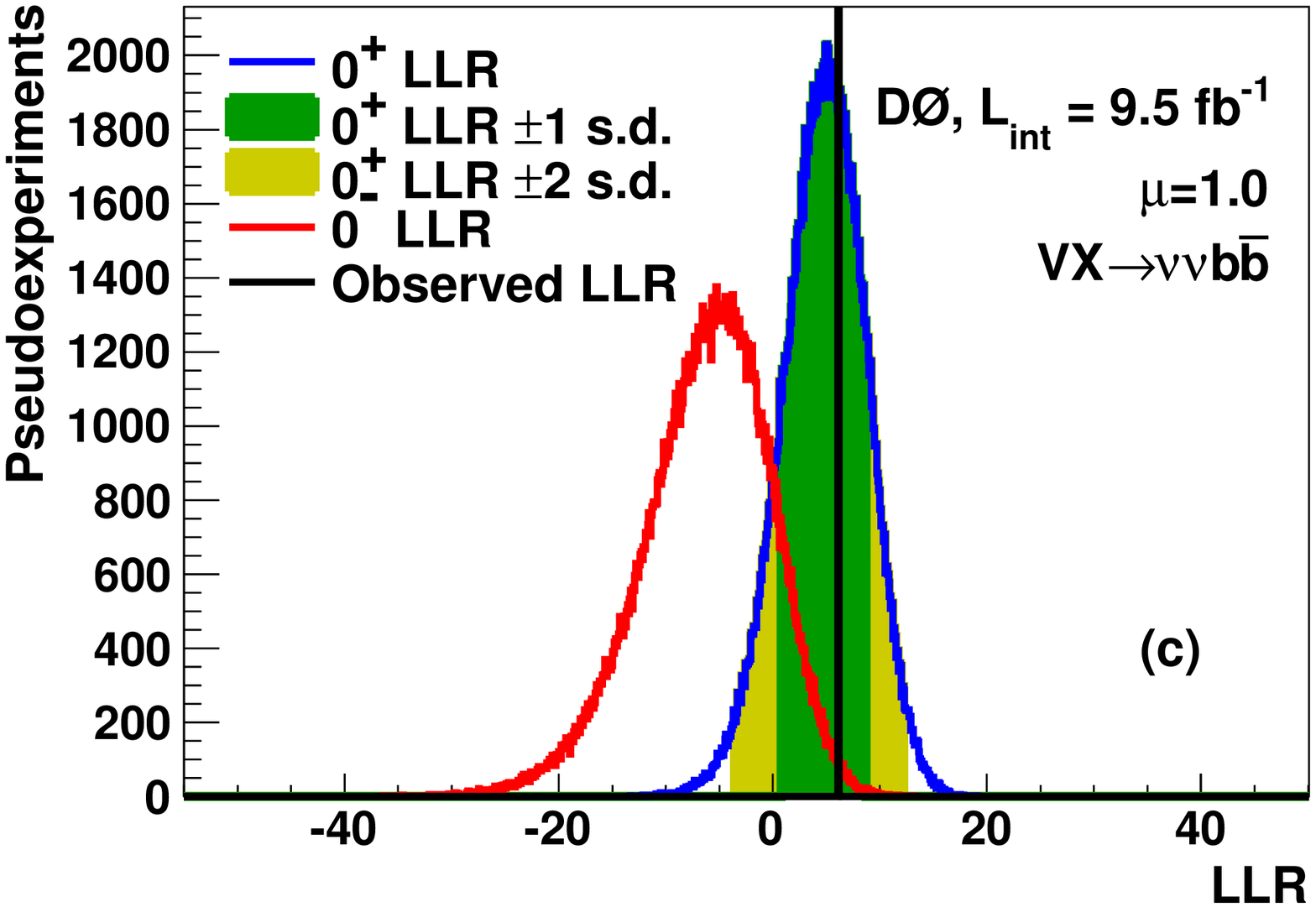}
\hspace{5 mm}
\includegraphics[width=0.47\textwidth]{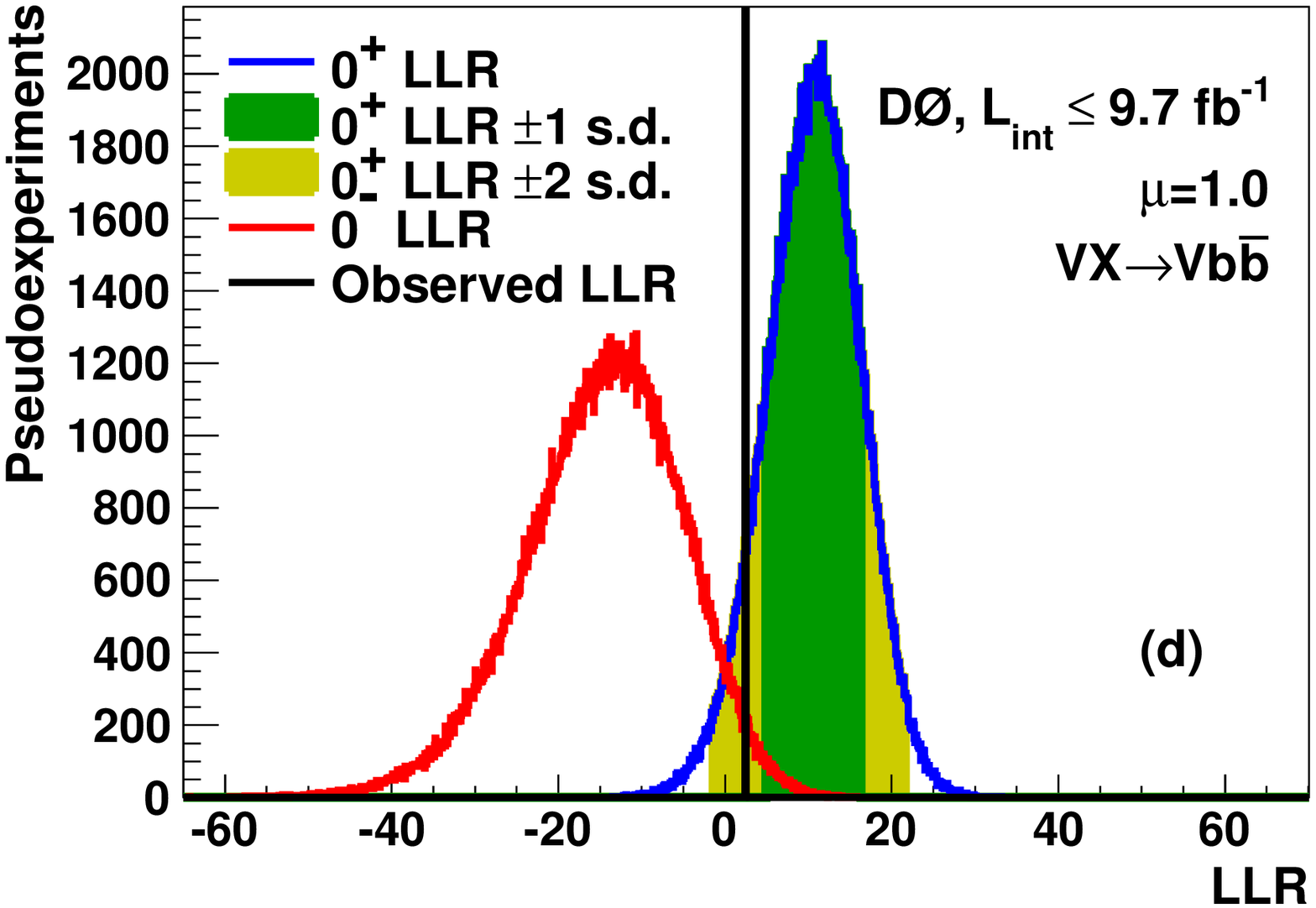}
\caption{LLR distributions comparing the \jpzp\ and the \jpzm\ hypotheses for the (a) \zhl\ analysis, 
(b) \whl\ analysis, (c) \zhv\ analysis, and (d) their combination.  
The \jpzp\ and \jpzm\ 
samples are normalized to the product of the SM cross section and branching fraction multiplied by $\mu=1.0$. The vertical solid line represents the observed 
LLR value, while the dark and light shaded areas represent 1 s.d.\ and 2 s.d.\ on the expectation from the null hypothesis 
$H_{0}$, respectively. Here $H_{0}$ is the SM \jpzp\ signal plus backgrounds.
For all signals, a mass of 125~GeV for the~$\!H$ or~$\!X$ boson is assumed.
\label{fig:llr_zm}}
\end{figure}

\begin{figure}[htbp]
\centering
\includegraphics[width=0.47\textwidth]{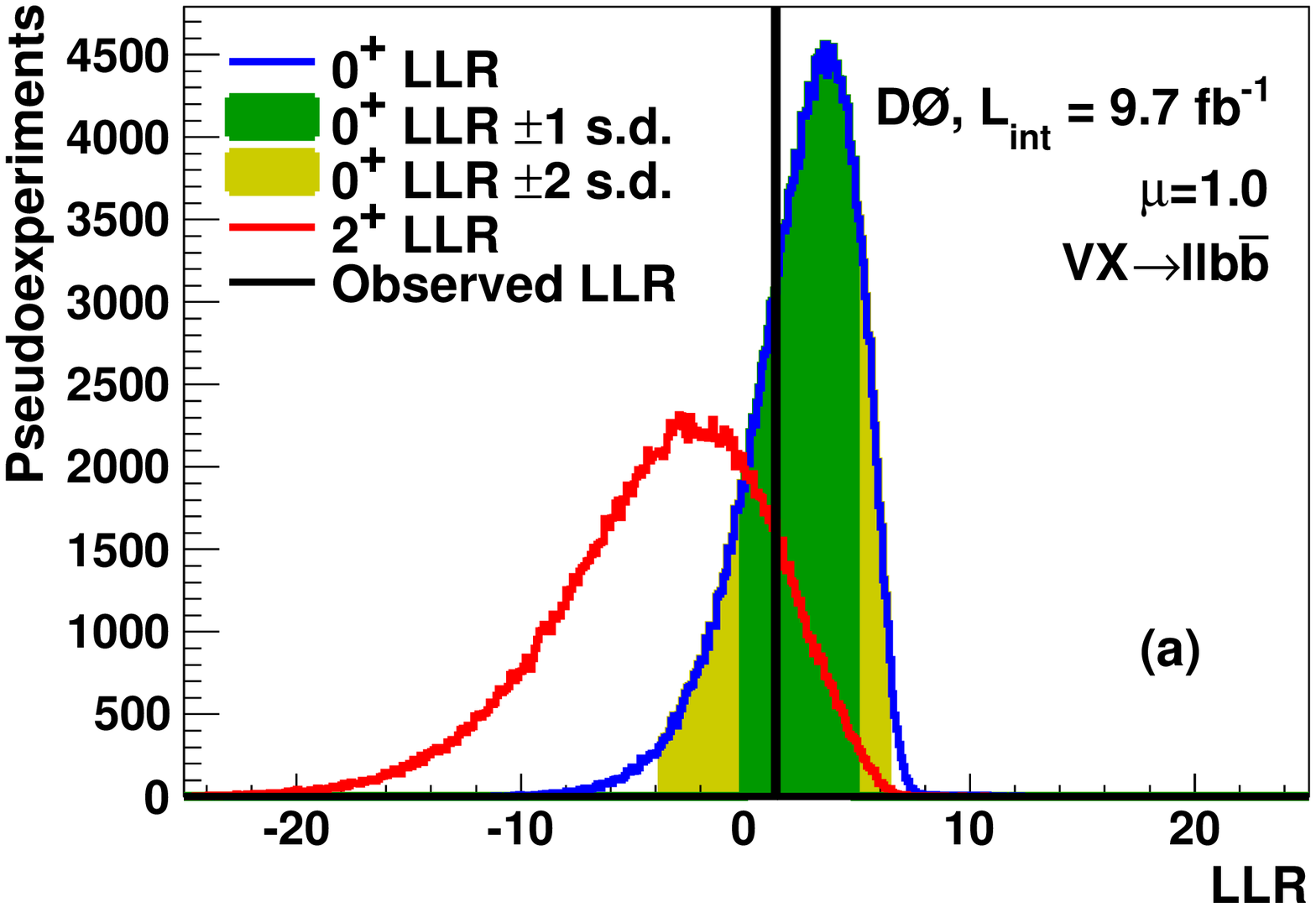}
\hspace{5 mm}
\includegraphics[width=0.47\textwidth]{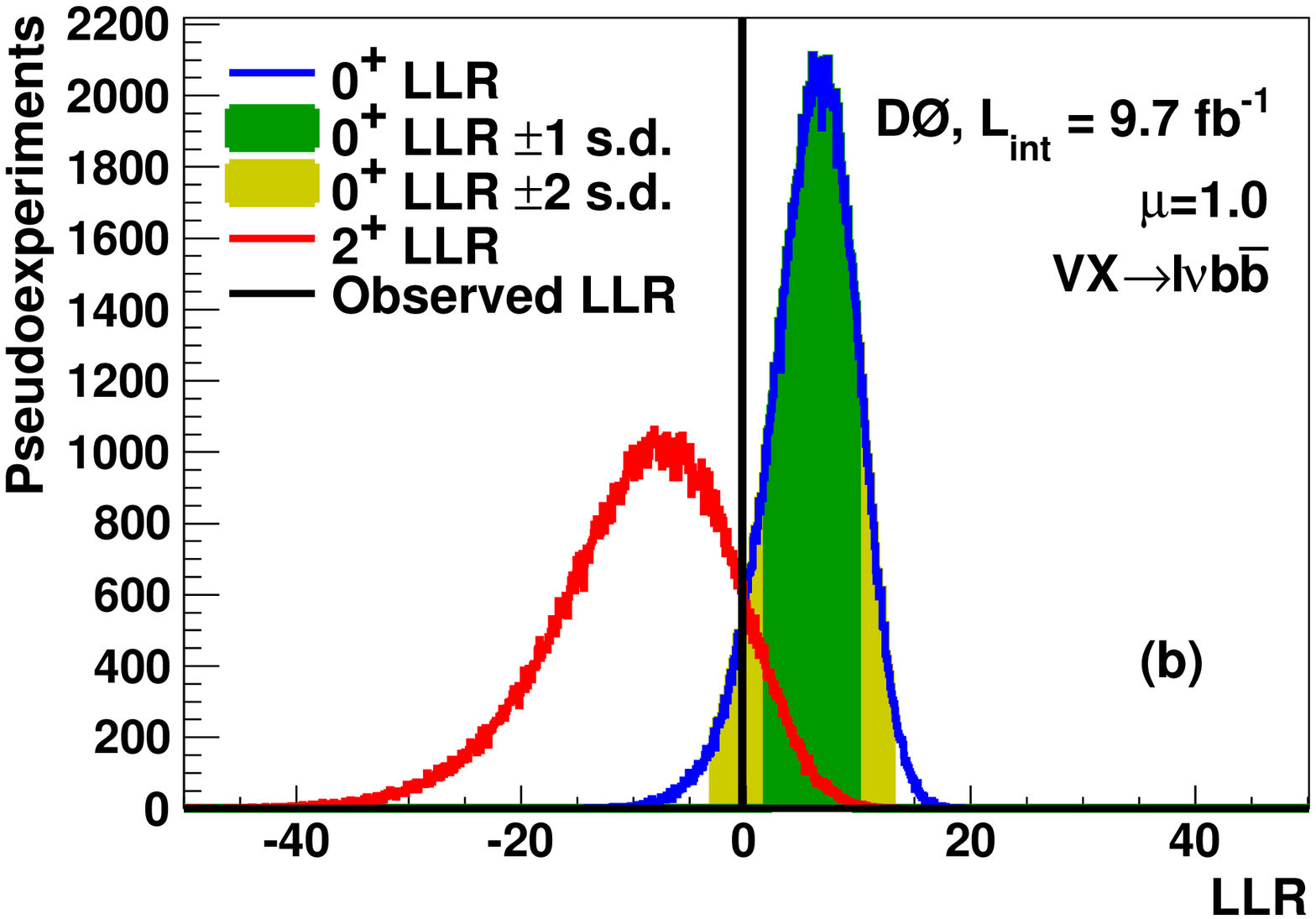}

\hspace{8 mm}

\includegraphics[width=0.47\textwidth]{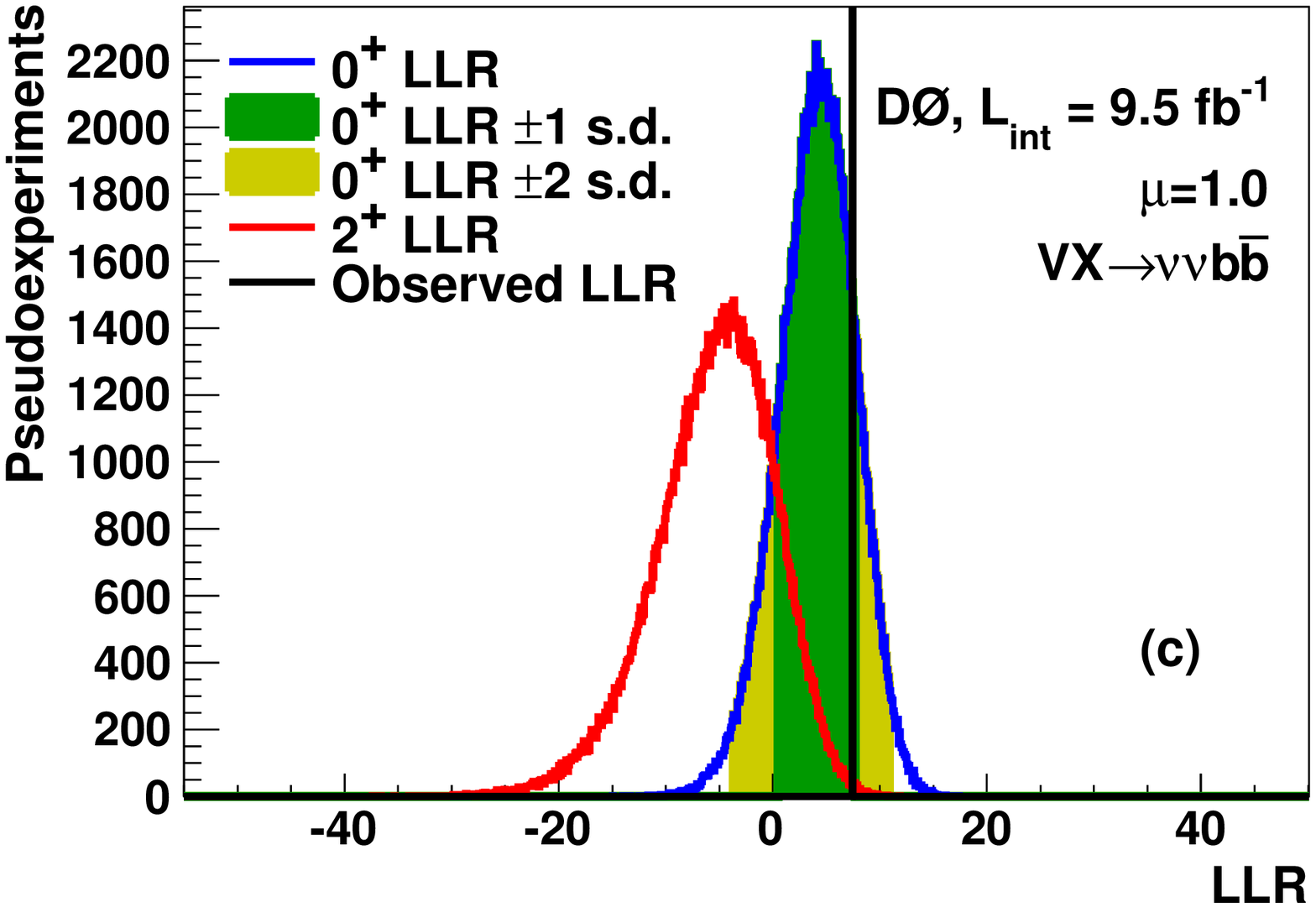}
\hspace{5 mm}
\includegraphics[width=0.47\textwidth]{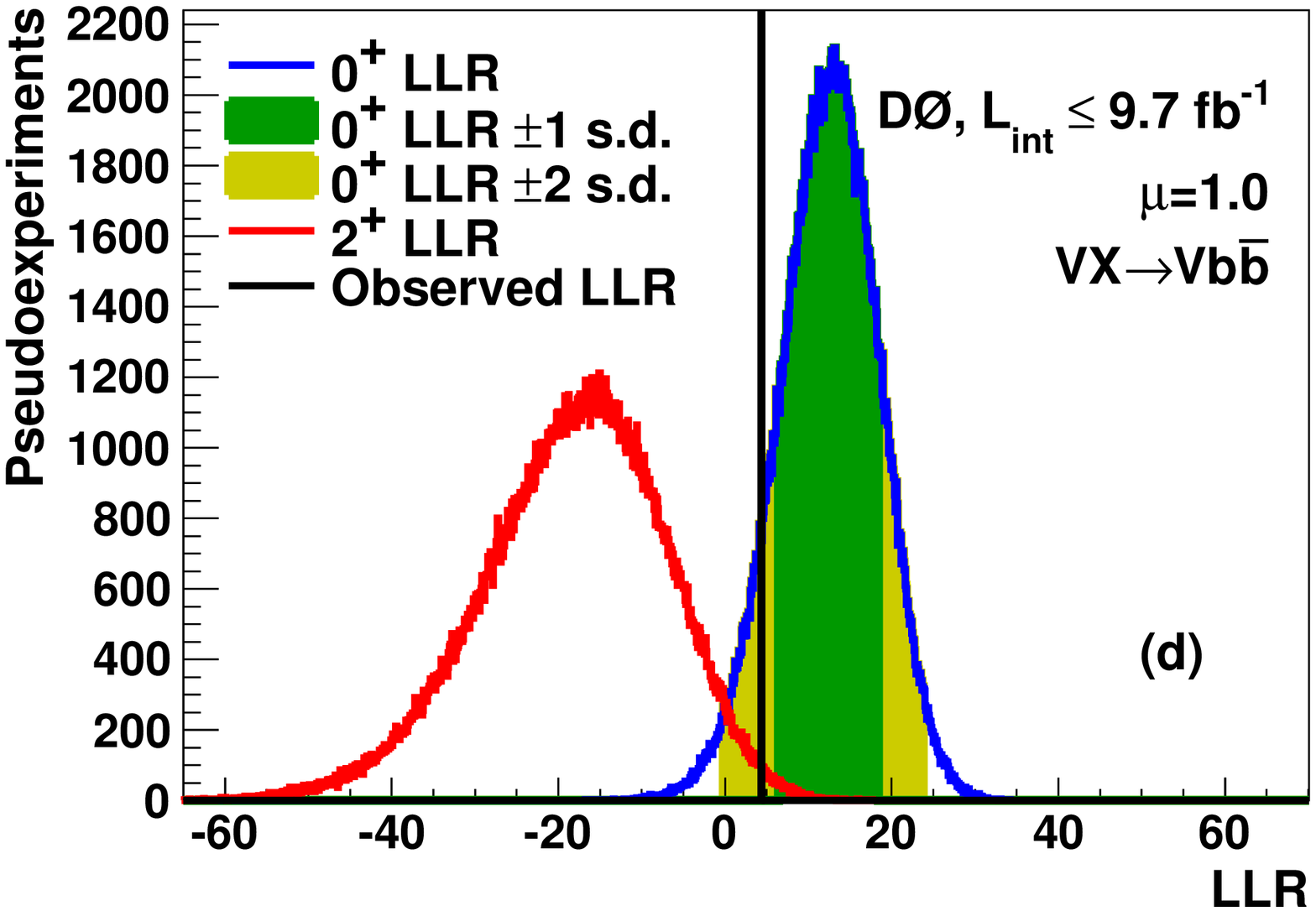}
\caption{LLR distributions comparing the \jpzp\ and the \jptp\ hypotheses for the (a) \zhl\ analysis, 
(b) \whl\ analysis, (c) \zhv\ analysis, and (d) their combination. 
The \jpzp\ and \jptp\ 
samples are normalized to the product of the SM cross section and branching fraction multiplied by $\mu=1.0$. The vertical solid line represents the observed 
LLR value, while the dark and light shaded areas represent 1 s.d.\ and 2 s.d.\ on the expectation from the null hypothesis 
$H_{0}$, respectively. Here $H_{0}$ is the SM \jpzp\ signal plus backgrounds.
For all signals, a mass of 125~GeV for the~$\!H$ or~$\!X$ boson is assumed.
\label{fig:llr_tp}}
\end{figure}

\clearpage

\begin{table}[htp]
\begin{center}
\begin{tabular}{c|@{\hskip 0.5cm}l@{\hskip 0.5cm}l@{\hskip 0.5cm}l@{\hskip 0.5cm}l}
\hline
\hline
Analysis & $ZH\to\ell\ell b\bar{b}$ & $WH\to\ell\nu b\bar{b}$ & $ZH\to\nu\nu b\bar{b}$ & Combined \T\B \\
\hline
 & \multicolumn{4}{c}{\jpzm\ vs. \jpzp} \T\B \\
\hline
\clzm\ Expected  & 0.075 & 0.030 & 0.016 & 0.0007 \T \\
\clzm\ Observed  & 0.126 & 0.351 & 0.007 & 0.022 \\
\clzp\ Expected  & 0.500   & 0.500   & 0.500   & 0.500 \\
\clzp\ Observed  & 0.646 & 0.965 & 0.367 & 0.918 \\
$1-CL_{s}$ Expected & 0.850 (1.04 s.d.) & 0.941 (1.56 s.d.) & 0.969 (1.87 s.d.) & 0.9986 (3.00 s.d.)\\ 
$1-CL_{s}$ Observed & 0.805 (0.86 s.d.) & 0.637 (0.35 s.d.) & 0.981 (2.07 s.d.) & 0.976 (1.98 s.d.)\B \\
\hline
& \multicolumn{4}{c}{\jptp\ vs. \jpzp} \T\B \\

\hline
\cltp\ Expected & 0.064  & 0.009 & 0.023 & 0.0003 \T \\
\cltp\ Observed & 0.134 & 0.114 & 0.002 & 0.009 \\
\clzp\ Expected & 0.500 & 0.500 & 0.500 & 0.500 \\
\clzp\ Observed  & 0.702 & 0.932 & 0.173 & 0.906 \\
$1-CL_{s}$ Expected & 0.872 (1.14 s.d.) & 0.982 (2.09 s.d.) & 0.953 (1.68 s.d.) & 0.9994 (3.22 s.d.)\\ 
$1-CL_{s}$ Observed & 0.810 (0.88 s.d.) & 0.878 (1.16 s.d.) & 0.987 (2.23 s.d.) & 0.990 (2.34 s.d.)\B \\
\hline
\hline
\end{tabular}
\caption{Expected and observed $CL_{H_{x}}$ and 
$1-CL_{s}$ values for \jpzm\ and \jptp$\,$\textit{VX} associated production, assuming signal cross sections equal to the
125~GeV SM Higgs production cross section multiplied by $\mu=1.0$. The null hypothesis is taken to be the sum 
of the SM Higgs 
boson signal and background production.\label{table:cls}}
\end{center}
\end{table}

\clearpage 

\begin{table}[htp]
\begin{center}
\begin{tabular}{c|@{\hskip 0.5cm}l@{\hskip 0.5cm}l@{\hskip 0.5cm}l@{\hskip 0.5cm}l}
\hline
\hline
Analysis & $ZH\to\ell\ell b\bar{b}$ & $WH\to\ell\nu b\bar{b}$ & $ZH\to\nu\nu b\bar{b}$ & Combined \T\B \\
\hline
 & \multicolumn{4}{c}{\jpzm\ vs. \jpzp} \T\B \\
\hline
\clzm\ Expected & 0.046 & 0.012 & 0.005 & $<$0.0001 \T \\
\clzm\ Observed & 0.072 & 0.245 & 0.0006 & 0.005 \\
\clzp\ Expected & 0.500   & 0.500   & 0.500   & 0.500 \\
\clzp\ Observed & 0.615 & 0.971 & 0.215 & 0.922 \\
$1-CL_{s}$ Expected & 0.908 (1.33 s.d.) & 0.975 (1.96 s.d.) & 0.989 (2.31 s.d.) & 0.9998 (3.60 s.d.)\\
$1-CL_{s}$ Observed & 0.883 (1.19 s.d.) & 0.747 (0.67 s.d.) & 0.997 (2.78 s.d.) & 0.995 (2.56 s.d.)\B \\
\hline
& \multicolumn{4}{c}{\jptp\ vs. \jpzp} \T\B \\
\hline
\cltp\ Expected & 0.037 & 0.003 & 0.009 & $<$0.0001 \T \\
\cltp\ Observed & 0.078 & 0.056 & 0.003 & 0.002 \\
\clzp\ Expected & 0.500 & 0.500 & 0.500 & 0.500 \\
\clzp\ Observed & 0.679 & 0.937 & 0.363 & 0.911 \\
$1-CL_{s}$ Expected & 0.925 (1.44 s.d.) & 0.995 (2.56 s.d.) & 0.983 (2.11 s.d.) & 0.9999 (3.86 s.d.)\\
$1-CL_{s}$ Observed & 0.885 (1.20 s.d.) & 0.941 (1.56 s.d.) & 0.991 (2.35 s.d.) & 0.998 (2.91 s.d.)\B \\
\hline
\hline
\end{tabular}
\caption{Expected and observed $CL_{H_{x}}$ and 
$1-CL_{s}$ values for \jpzm\ and \jptp$\,$\textit{VX} associated production, assuming signal cross sections equal to the
125~GeV SM Higgs production cross section multiplied by $\mu=1.23$. The null hypothesis is taken to be the sum 
of the SM Higgs 
boson signal and background production.\label{table:cls_1.23}}
\end{center}
\end{table}

\clearpage

\begin{figure}[htp]
\centering
\includegraphics[width=0.80\textwidth]{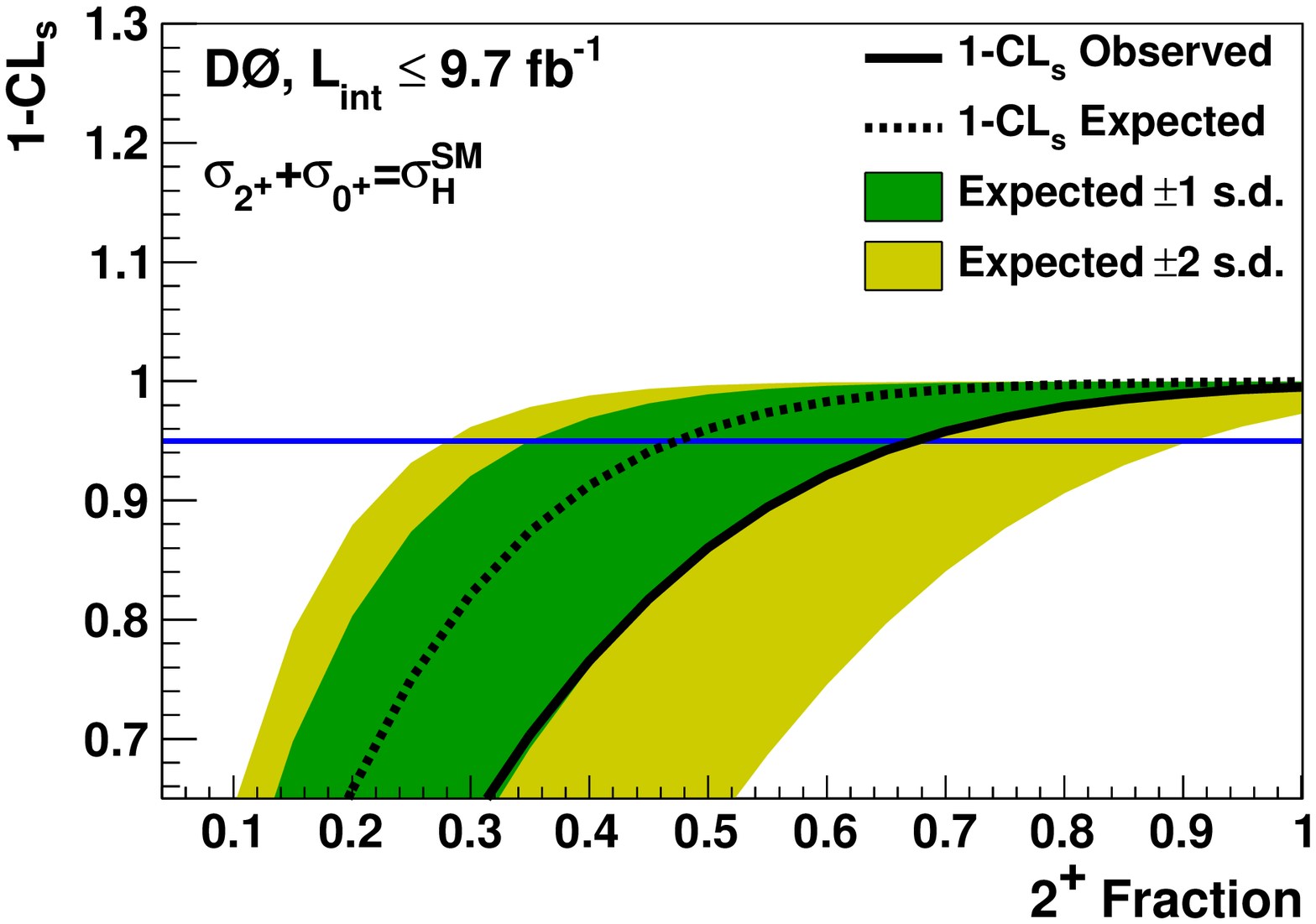}
\caption{\label{fig:sig_frac_jptp}(color online) $1-CL_{s}$ as a function of the \jptp\ signal fraction 
\ftplusm\ for $\mu=1.0$ for all analyses combined. The horizontal solid line corresponds to the 95\%\ CL exclusion. The 
dark and light shaded regions represent the expected 1 and 2 s.d.\ fluctuations of the \jpzp\ hypothesis.}
\end{figure}

\begin{figure}[htp]
\centering
\includegraphics[width=0.80\textwidth]{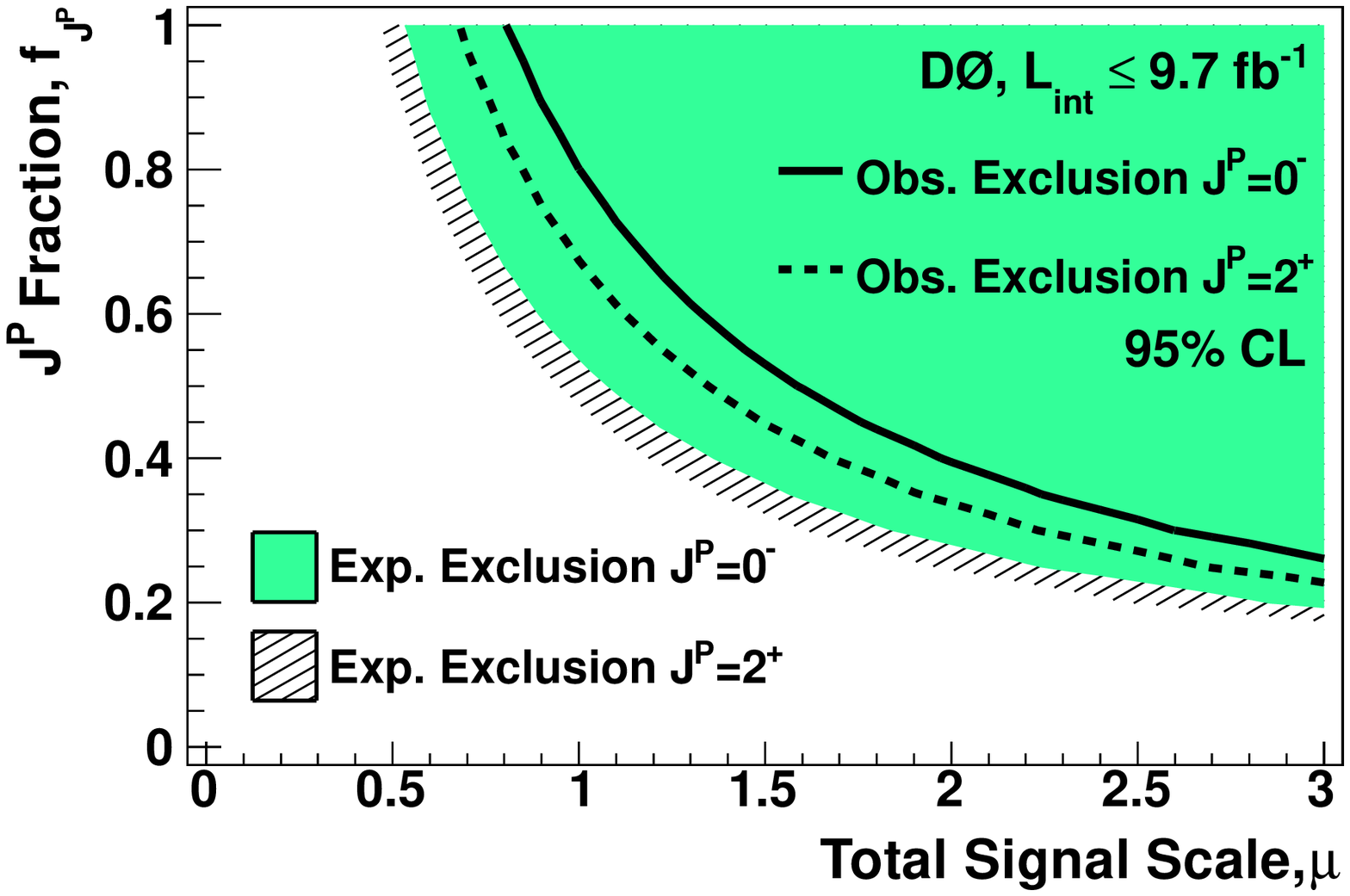}
\caption{\label{fig:fracrate}(color online) The  expected 95\%\ CL exclusion (shaded area) and observed 95\%\ CL exclusion (solid line) as functions of the \jpzm\ signal fraction \fzminusm\ 
and the total signal strength in units of the SM Higgs cross section multiplied by the branching ratio. As functions 
of the \jptp\ signal fraction \ftplusm\ and the total signal strength, the expected and observed exclusions are shown as the hatched area and dashed line, 
respectively.}
\end{figure}

\FloatBarrier

\FloatBarrier
\end{widetext}

\end{document}